\documentclass[preprint,floats,aps,epsfig,nofootinbib,amssymb,showpacs]{revtex4}
\usepackage{color}
\usepackage{graphicx}
\usepackage{tabularx}
\usepackage{amsmath}
\usepackage{amssymb}
\usepackage{datetime}
\usepackage[left]{lineno}
\usepackage{txfonts}
\usepackage{dcolumn}
\usepackage{bm}

\begin{document}

\title{Searching for Dark Matter Signals in the Left-Right Symmetric Gauge Model with $CP$ Symmetry}

\author{Wan-Lei Guo}
\email[Email: ]{guowl@itp.ac.cn}

\author{Yue-Liang Wu}
\email[Email: ]{ylwu@itp.ac.cn}

\author{Yu-Feng Zhou}
\email[Email: ]{yfzhou@itp.ac.cn}

\affiliation{ Kavli Institute for Theoretical Physics China, \\
Key Laboratory of Frontiers in Theoretical Physics, \\
Institute of Theoretical Physics, Chinese Academy of Science,
Beijing 100190, China}

\begin{abstract}
  We investigate  singlet scalar dark matter (DM) candidate in a left-right
  symmetric gauge model with two Higgs bidoublets (2HBDM) in which the
  stabilization of the DM particle is induced by the discrete symmetries $P$
  and $CP$. According to the observed DM abundance, we predict the DM direct
  and indirect detection cross sections for the DM mass range from 10 GeV to
  500 GeV. We show that the DM indirect detection cross section is not
  sensitive to the light Higgs mixing and Yukawa couplings except the
  resonance regions. The predicted spin-independent DM-nucleon elastic
  scattering cross section is found to be significantly dependent on the above
  two factors. Our results show that the future DM direct search experiments
  can cover the most parts of the allowed parameter space. The PAMELA antiproton
  data can only exclude two very narrow regions in the 2HBDM. It is very
  difficult to detect the DM direct or indirect signals in the resonance
  regions due to the Breit-Wigner resonance effect.

\end{abstract}


\maketitle

\section{Introduction}

The existence of dark matter (DM) is by now well established from
astrophysical observations \cite{DM}. Together with the recent WMAP
results, the cosmological observations have shown that the present
Universe consists of about 73\% dark energy, 23\% dark matter, and
4\% baryonic matter~\cite{WMAP7}.  In the standard model (SM) of
particle physics, there is no cold DM candidate. Therefore, one has
to extend the SM to account for the existence of DM.  The DM
candidate is often accompanied by some discrete symmetries to keep
it stable, such as the R parity in supersymmetric (SUSY) models and
KK parity in universal extra dimension models. Although the discrete
symmetries are necessary for the DM stability, they may be
introduced from different motivations~\cite{DM}.

In the left-right (LR) symmetric gauge model
\cite{LRmodel,Beall:1981ze,Deshpande:1990ip} with spontaneous $CP$
violation (SCPV), the $P$ and $CP$ symmetries are exact before the
spontaneous symmetry breaking (SSB). In this case, it is possible
that the discrete symmetries $P$ and $CP$ strongly constrain the
scalar sector of the model and naturally give stable DM candidates.
This possibility has not been emphasized in the literature, due to
the fact that most of the popular models such as SM and SUSY violate
$P$ maximally. In Ref. \cite{Guo:2008si}, we have shown that the $P$
and $CP$ symmetries can give a stable DM candidate in an extension
of a left-right symmetric gauge model with a singlet scalar field $S
= (S_\sigma + i S_D)/\sqrt{2}$. In this model, the $CP$ odd particle
$S_D$ is stable even after the SSB, provided that it does not
develop vacuum expectation value (VEV).

Without large fine-tuning, it is difficult to have a successful SCPV
in the minimal left-right symmetric gauge model with only one Higgs
bidoublet (1HBDM) \cite{Deshpande:1990ip,Masiero:1981zd}.  This is
because in the decoupling limit the predicted $CP$ violating
quantity $\sin2\beta\sim 0.1$ with $\beta$ being a $CP$ phase angle
in the Cabibbo-Kobayashi-Maskawa (CKM) matrix is far below the
experimentally measured value of $\sin2\beta=0.671\pm 0.024$ from
the two B-factories \cite{Ball:1999mb}. In addition, the 1HBDM is
also subject to strong phenomenological constraints from low energy
flavor changing neutral current (FCNC) processes, especially the
neutral kaon mixing which pushes the masses of the right-handed
gauge bosons and some neutral Higgs bosons much above the TeV scale
\cite{Mohapatra:1983ae}.  Motivated by the requirement of both
spontaneous $P$ and $CP$ violations, we have considered the
left-right symmetric gauge model with two Higgs bidoublets (2HBDM)
\cite{Wu:2007kt}. In the 2HBDM, the additional Higgs bidoublet
modifies the Higgs potential so that the fine-tuning problem in the
SCPV can be avoided, and the bounds from the FCNC processes can be
relaxed. The extra Higgs bidoublet may also change the interferences
among different contributions in the neutral meson mixings, and
lower the bounds for the right-handed gauge boson masses not to be
much higher than the TeV scale~\cite{Wu:2007kt}.  Such a
right-handed gauge boson can be searched at the LHC using the
angular distributions of top quarks and the leptons from top quark
decays \cite{Gopalakrishna:2010xm}.

In Ref. \cite{Guo:2008si}, we have shown that the discrete
symmetries $P$ and $CP$ can be used to stabilize the DM candidate
$S_D$ in the 1HBDM and 2HBDM with the SCPV. Using the observed DM
abundance, we can constrain the parameter space and predict the
spin-independent (SI) DM-nucleon elastic scattering cross section.
For simplicity, we have only considered the case with no mixing
among light neutral Higgs bosons in the 2HBDM and the dark matter is
heavy. In this paper, we shall demonstrate in detail the mixing
effect on the DM direct detection. Notice that several new DM
annihilation channels can be derived, namely two DM particles may
annihilate into a gauge boson and a Higgs boson. On the other hand,
we are going to extend the DM mass range from $200 \;{\rm GeV} \leq
m_D \leq 500$ GeV to $10 \;{\rm GeV} \leq m_D \leq 500$ GeV. As a
consequence, one will meet several resonances in the 2HBDM.
Therefore we shall consider the Breit-Wigner resonance effect for
the determination of the DM relic density \cite{BW}. In addition, we
will also consider the DM indirect search in the 1HBDM and 2HBDM.
The paper is organized as follows: In Section. \ref{Model}, we
outline the main features of the 1HBDM and 2HBDM with a singlet
scalar. In Sec. \ref{Sec1HBDM} and Sec. \ref{Sec2HBDM}, we discuss
the parameter space, the DM direct search and the DM indirect search
in the 1HBDM and 2HBDM, respectively. Some conclusions are given in
Sec. \ref{Conclusion}.

\section{The left-right symmetric gauge model with a singlet scalar}\label{Model}

We begin with a brief review of the 2HBDM described in Ref.
\cite{Wu:2007kt}.  The model is a simple extension to the 1HBDM,
which is based on the gauge group $SU(2)_L\otimes SU(2)_R\otimes
U(1)_{B-L}$. The left- and right-handed fermions belong to $SU(2)_L$
and $SU(2)_R$ doublets, respectively. The Higgs sector  contains two
Higgs bidoublets $\phi$ (2,$2^{*}$,0), $\chi$ (2,$2^{*}$,0) and a
left(right)-handed Higgs triplet $\Delta_{L(R)}$ (3(1),1(3),2) with
the following flavor contents
\begin{eqnarray}
\phi  = \left ( \begin{matrix} \phi_1^0 & \phi_2^+ \cr \phi_1^- &
\phi_2^0 \cr  \end{matrix} \right ) , \; \chi  = \left (
\begin{matrix} \chi_1^0 & \chi_2^+ \cr \chi_1^- & \chi_2^0 \cr
\end{matrix} \right ) , \; \Delta_{L,R}  = \left (
\begin{matrix} \delta_{L,R}^+/\sqrt{2} & \delta_{L,R}^{++} \cr
\delta_{L,R}^{0} & -\delta_{L,R}^{+}/\sqrt{2} \cr \end{matrix}
\right ) . \label{Higgscomponent}
\end{eqnarray}
The introduction of Higgs bidoublets $\phi$ and $\chi$ can account
for the electroweak symmetry breaking and overcome the fine-tuning
problem in generating the SCPV in the 1HBDM. Meanwhile it also
relaxes the severe low energy phenomenological constraints
\cite{Wu:2007kt}. Motivated by the spontaneous $P$ and $CP$
violations, we require $P$ and $CP$ invariance of the Lagrangian,
which strongly restricts the structure of the Higgs potential. The
most general potential containing only the $\phi$ and $\Delta_{L,R}$
fields is given by
\begin{eqnarray}
\mathcal{V}_{\phi\Delta} & = & -\mu_{1}^{2}\rm{Tr}(\phi^{\dagger}\phi)-\mu_{2}^{2}[\rm{Tr}(\tilde{\phi}^{\dagger}\phi)+\rm{Tr}(\tilde{\phi}\phi^{\dagger})]-\mu_{3}^{2}[\rm{Tr}(\Delta_{L}\Delta_{L}^{\dagger})+\rm{Tr}(\Delta_{R}\Delta_{R}^{\dagger})]\nonumber\\
 &  & +\lambda_{1}[\rm{Tr}(\phi^{\dagger}\phi)]^{2}+\lambda_{2}\{[\rm{Tr}(\tilde{\phi}^{\dagger}\phi)]^{2}+[\rm{Tr}(\tilde{\phi}\phi^{\dagger})]^{2}\}+\lambda_{3}[\rm{Tr}(\tilde{\phi}^{\dagger}\phi)\rm{Tr}(\tilde{\phi}\phi^{\dagger})]\nonumber\\
 &  & +\lambda_{4}\{\rm{Tr}(\phi^{\dagger}\phi)[\rm{Tr}(\tilde{\phi}^{\dagger}\phi)+\rm{Tr}(\tilde{\phi}\phi^{\dagger})]\}\nonumber\\
 &  & +\rho_{1}\{[\rm{Tr}(\Delta_{L}\Delta_{L}^{\dagger})]^{2}+[\rm{Tr}(\Delta_{R}\Delta_{R}^{\dagger})]^{2}\}+\rho_{2}[\rm{Tr}(\Delta_{L}\Delta_{L})\rm{Tr}(\Delta_{L}^{\dagger}\Delta_{L}^{\dagger})+\rm{Tr}(\Delta_{R}\Delta_{R})\rm{Tr}(\Delta_{R}^{\dagger}\Delta_{R}^{\dagger})]\nonumber\\
 &  & +\rho_{3}[\rm{Tr}(\Delta_{L}\Delta_{L}^{\dagger})\rm{Tr}(\Delta_{R}\Delta_{R}^{\dagger})]+\rho_{4}[\rm{Tr}(\Delta_{L}\Delta_{L})\rm{Tr}(\Delta_{R}^{\dagger}\Delta_{R}^{\dagger})+\rm{Tr}(\Delta_{L}^{\dagger}\Delta_{L}^{\dagger})\rm{Tr}(\Delta_{R}\Delta_{R})]\nonumber\\
 &  & +\alpha_{1}\rm{Tr}(\phi^{\dagger}\phi)[\rm{Tr}(\Delta_{L}\Delta_{L}^{\dagger})+\rm{Tr}(\Delta_{R}\Delta_{R}^{\dagger})]
          +\alpha_{2}\rm{Tr}[ (\tilde{\phi}^{\dagger}\phi)+(\tilde{\phi}\phi^{\dagger})]\rm{Tr}[(\Delta_{L}\Delta_{L}^{\dagger})+(\Delta_{R}\Delta_{R}^{\dagger})]\nonumber\\
 &  & +\alpha_{3}[\rm{Tr}(\phi\phi^{\dagger}\Delta_{L}\Delta_{L}^{\dagger})+\rm{Tr}(\phi^{\dagger}\phi\Delta_{R}\Delta_{R}^{\dagger})]\nonumber\\
 &  & +\beta_{1}[\rm{Tr}(\phi\Delta_{R}\phi^{\dagger}\Delta_{L}^{\dagger})+\rm{Tr}(\phi^{\dagger}\Delta_{L}\phi\Delta_{R}^{\dagger})]+\beta_{2}[\rm{Tr}(\tilde{\phi}\Delta_{R}\phi^{\dagger}\Delta_{L}^{\dagger})+\rm{Tr}(\tilde{\phi}^{\dagger}\Delta_{L}\phi\Delta_{R}^{\dagger})]\nonumber\\
 &  & +\beta_{3}[\rm{Tr}(\phi\Delta_{R}\tilde{\phi}^{\dagger}\Delta_{L}^{\dagger})+\rm{Tr}(\phi^{\dagger}\Delta_{L}\tilde{\phi}\Delta_{R}^{\dagger})] ,
\label{Vphidelta}
\end{eqnarray}
where the coefficients $\mu_i$, $\lambda_i$, $\rho_i$, $\alpha_i$
and $\beta_i$ in the potential are all real as all the terms are
self-Hermitian. The Higgs potential $\mathcal{V}_{\chi\Delta}$
involving $\chi$ field can be obtained by the replacement
$\chi\leftrightarrow \phi$ in Eq. (\ref{Vphidelta}). The mixing term
$\mathcal{V}_{\chi\phi\Delta}$ can be obtained by replacing one of
$\phi$ by $\chi$ in all the possible ways in Eq. (\ref{Vphidelta}).
In order to simplify the discussion, we shall first consider the
1HBDM which already contains the main features of the complete
model. Then we postpone the discussions on the $\chi$ contributions
into Section \ref{Sec2HBDM}.

After the SSB, the Higgs multiplets obtain  nonzero VEVs
\begin{equation}
\langle\phi_{1,2}^0\rangle=\frac{\kappa_{1,2}}{\sqrt2} \; \mbox{ and
}\; \langle\delta_{L,R}^0\rangle=\frac{v_{L,R}}{\sqrt2}\;,
\end{equation}
where $\kappa_1$, $\kappa_2$, $v_L$ and $v_R$ are in general
complex, and $ \kappa \equiv \sqrt{|\kappa_1|^2 + |\kappa_2|^2}
\approx 246$ GeV represents the electroweak symmetry breaking scale.
Due to the freedom of gauge symmetry transformation, one can take
$\kappa_1$ and $v_R$ to be real. To avoid the fine-tuning problem of
fermion masses, we require $v_L \simeq 0$ and $\kappa_2 \ll
\kappa_1$. The value of $v_R$ sets the scale of left-right symmetry
breaking which is directly linked to the right-handed gauge boson
masses. $v_R$  is subjected to strong constraints from the $K$, $B$
meson mixings \cite{Mohapatra:1983ae,Ball:1999mb,Beall:1981ze} as
well as low energy electroweak interactions
\cite{Barenboim:1996nd,Langacker:1989xa}. The kaon mass difference
and the indirect $CP$ violation quantity $\epsilon_K$ set a bound
for $v_R$ around $10$ TeV \cite{Barenboim:1996nd,Pospelov:1996fq}.

\begin{table}[htb]
\begin{center}
\begin{tabular}{|c|c|c||c|c|c||c|c|c|}
\hline   &  $P$  &   $CP$  & &\;\; $P$\;\; & \;$CP$\; &  & \;\; $P$\;\; & \;$CP$\; \\
\hline $\phi$ \;   &  $\phi^\dagger$ \; & $\phi^*$ & $S + S^*$  & + & + & $S - S^*$  & + & - \\
\hline  $\chi$ \;   &  $\chi^\dagger$ \; &$\chi^*$ & $S S^*$ & + & + &  Tr($\phi^{\dag} \phi$) & + & + \\
\hline $\Delta_{L(R)}$   & $\Delta_{R(L)}$  & $\Delta_{L(R)}^*$ & Tr($\phi^{\dag} \tilde{\phi}+ \tilde{\phi}^{\dag} \phi $)  & + & + & Tr($\phi^{\dag} \tilde{\phi}- \tilde{\phi}^{\dag} \phi $)  & - & - \\
\hline $S$  &  $S$ &   $S^*$ & Tr($\Delta_L^{\dag} \Delta_L +
\Delta_R^{\dag} \Delta_R$)  & + & + & Tr($\Delta_L^{\dag} \Delta_L -
\Delta_R^{\dag} \Delta_R$)  &
- & + \\
\hline
\end{tabular}
\end{center}
\vspace{-0.2cm} \caption{The $P$ and $CP$ transformation properties
of the Higgs particles and their gauge-invariant combinations. The
``+" and ``-" denote even and odd, respectively.  } \label{PCP}
\end{table}

In our model, the $P$ and $CP$ symmetries have been required to be
exactly conserved before the SSB, thus the discrete symmetries $P$
and $CP$ can be used to stabilize the DM candidate. In the framework
of 2HBDM with a complex singlet scalar $S = (S_\sigma+i
S_D)/\sqrt2$, we have considered this possibility in Ref.
\cite{Guo:2008si}. The $P$ and $CP$ transformation properties  of
the Higgs particles and their gauge-invariant combinations  have
been shown in Table \ref{PCP}. It is clear that the odd powers of
$(S-S^*)$ are forbidden by the $P$ and $CP$ symmetries. Therefore
$S_D$ is a stable particle and can be the DM candidate when the VEV
$v_{\sigma}/\sqrt{2}$ of $S$ is real. Although $P$ and $CP$ are both
broken after the SSB, there is a $CP$ type $Z_2$ discrete symmetry
on $S_D$ remaining in the singlet sector. This discrete symmetry is
induced from the original  $CP$ symmetry. We have checked that the
$P$ and $CP$ transformation rules for $S$ defined in Table \ref{PCP}
is actually the only possible way for the implementation of the DM
candidate.

For the annihilation cross section of approximately weak strength,
we expect that the DM mass is in the range of a few GeV and a few
hundred GeV. However, the mass $m_D$ of $S_D$ is related to the LR
symmetry breaking scale  $v_R \sim 10$ TeV. To have a possible
light DM mass, we may consider an approximate global $U(1)$ symmetry
on $S$, i.e. $S\to e^{i\delta}S$. Then the $P$ and $CP$ invariant
Higgs potential involving the singlet $S$ is given by
\begin{eqnarray}
\mathcal{V}_{S}& = & - \mu_D^2 SS^* + \lambda_D (SS^*)^2 +
\sum_{i=1}^3 \lambda_{i,D} SS^*  O_i - \frac{ m_D^2}{4} (S -
S^*)^2\;,
\end{eqnarray}
where $O_1={\rm{Tr}}(\phi^{\dag}\phi)$,
$O_2={\rm{Tr}}(\phi^{\dag}\tilde\phi+\tilde\phi^{\dag}\phi)$ and
$O_3={\rm{Tr}}(\Delta_L^{\dag}\Delta_L+\Delta_R^{\dag}\Delta_R)$.
Only the last term explicitly violates $U(1)$ symmetry. After the
SSB, $S$ obtains a real VEV $v_\sigma/\sqrt{2}$. Then one can
straightly derive
\begin{eqnarray}
\mathcal{V}_{S}  =  \frac{\lambda_{D}}{4} [(S_{\sigma}^2 + 2
 v_\sigma S_\sigma  + S_D^2)^2  -  v_\sigma ^4]  + \sum_{i=1}^3 \frac{\lambda_{i,D}}{2} (S_{\sigma}^2 + 2
 v_\sigma S_\sigma  + v_\sigma ^2 + S_D^2) (O_{i} - \langle O_i\rangle)
+ \frac{m_D^2}{2}  S_D^2 \,, \label{VS}
\end{eqnarray}
where we have used the minimization condition $\mu_{D}^2 =
\lambda_{D} v_{\sigma}^2 + \sum_{i} \lambda_{i,D}\langle O_i\rangle$
from the singlet $S_{\sigma}$ to eliminate the  parameter $\mu_D$.
The terms proportional to odd powers of $S_D$ are absent in Eq.
(\ref{VS}) which implies $S_D$ can only be produced by pairs. Notice
that the mass term of $S_D$ should be absent with an exact global
$U(1)$ symmetry.  As discussed in Ref. \cite{Guo:2008si}, the
explicit breaking of this $U(1)$ symmetry can explain the
naturalness of a light DM mass $m_D$, but it does not destroy the
stability of the DM candidate $S_D$.

\begin{table}[thb]
\begin{center}
\begin{tabular}{|l|l||l|l|}
\hline Particles  & Mass$^2$ & Particles  & Mass$^2$\\\hline \hline
$h^0 = \phi_1^{0 r}$ &  $m_{h^0}^2 = 2 \lambda_1 \kappa^2$ & $H_2^{\pm} = \phi_2^{\pm}$ &  $m_{H_2^\pm}^2 = \frac{1}{2} \alpha_3 v_R^2 $ \\
$H_1^0 = \phi_2^{0 r}$ &  $m_{H_1^0}^2 = \frac{1}{2} \alpha_3 v_R^2$ & $H_R^{\pm \pm}=\delta_R^{\pm \pm}$ &   $m_{H_R^{\pm \pm}}^2 = 2 \rho_2 v_R^2$ \\
$A_1^0 = - \phi_2^{0 i}$ &  $m_{A_1^0}^2 = \frac{1}{2} \alpha_3
v_R^2$
& $H_L^{\pm}=\delta_L^{\pm}$ &   $m_{H_L^{\pm}}^2 = \frac{1}{2} (\rho_3 - 2 \rho_1) v_R^2$ \\
$H_2^0 = \delta_R^{0 r}$ &  $m_{H_2^0}^2 = 2 \rho_1 v_R^2 $ &
$H_L^{\pm \pm}=\delta_L^{\pm \pm}$
& $m_{H_L^{\pm \pm}}^2 = \frac{1}{2} (\rho_3 - 2 \rho_1) v_R^2$ \\
$H_3^0=\delta_L^{0r}$&   $m_{H_3^0}^2 = \frac{1}{2} (\rho_3 - 2
\rho_1) v_R^2 $ & $A_L^0=\delta_L^{0i}$& $m_{A_L^0}^2 = \frac{1}{2}
(\rho_3 - 2 \rho_1) v_R^2 $\\ \hline \hline  $Z_1$ & $m_{Z_1}^2 =
m_{W_1}^2\sec^2\theta_W$
&$W_1^\pm = W_L^\pm$ & $m_{W_1}^2 = g^2\kappa^2/4$ \\
$Z_2$ & $m_{Z_2}^2 = \frac{g^2 v_R^2 \cos^2 \theta_W}{\cos 2
\theta_W}$ &$W_2^\pm = W_R^\pm$ & $m_{W_2}^2 = g^2 v_R^2/2$
\\\hline
\end{tabular}
\end{center}
\vspace{-0.2cm} \caption{The mass spectrum for the Higgs and  gauge
bosons in the left-right symmetric gauge model with one Higgs
bidoublet in the limit $v_L \simeq 0$ and $\kappa_2 \ll \kappa_1$.
$\phi^{0r}_i$ and $\phi^{0i}_i$ stand for real and imaginary
components of $\phi^{0}_i = (\phi^{0r}_i + i \phi^{0i}_i)/\sqrt{2}$,
respectively. The gauge boson $Z_1(W_1)$ corresponds to the $Z(W)$
boson in the SM. } \label{mass}
\end{table}

The terms $2 v_\sigma S_\sigma O_{i}$ in Eq. (\ref{VS}) indicate
that $S_\sigma$ will mix with the Higgs bosons ${\phi_1^{0r}}$,
${\phi_2^{0r}}$, ${\delta_L^{0r}}$ and ${\delta_R^{0r}}$. The
relevant mass matrix elements are given by
\begin{eqnarray}
M^2_{\sigma} = 2\lambda_{D} v_{\sigma}^2 \, ,
M^2_{\sigma\phi_1^{0r}} = \lambda_{1,D}  \kappa  v_{\sigma} \, ,
M^2_{\sigma\phi_2^{0r}} = 2 \lambda_{2,D} \kappa v_{\sigma}\, ,
M^2_{\sigma\delta_L^{0r}} = \lambda_{3,D} v_{\sigma} v_L \, ,
M^2_{\sigma\delta_R^{0r}} = \lambda_{3,D} v_{\sigma} v_R\;.
\label{msigma}
\end{eqnarray}
For simplicity here we require $v_\sigma > v_R \sim 10 \; {\rm TeV}
\gg \kappa$ which means  the mixing angles between $S_\sigma$ and
the above four neutral Higgs bosons are  small. The terms
$v_\sigma^2 O_{i}$ in Eq. (\ref{VS}) do not change the minimization
condition forms for $\phi$ and $\Delta_{L(R)}$. This is because
these terms only change the overall coefficients $\mu_1$, $\mu_2$
and $\mu_3$ in Eq. (\ref{Vphidelta}). Hence the mass matrixes of the
Higgs multiplets $\phi$ and $\Delta_{L,R}$ remain the same as that
in the 1HBDM in Refs. \cite{Deshpande:1990ip,Duka:1999uc}, which
also indicates that the additional potential term $\mathcal{V}_S$ in
Eq. (\ref{VS}) does not help in resolving the fine-tuning problem.
Due to $v_L \simeq 0$ and $\kappa_2 \ll \kappa_1$, the mass
eigenstates for the Higgs bidoublet and triplets approximately
coincide with the corresponding flavor eigenstates. The mass
spectrum for the Higgs and gauge bosons is listed in Table
\ref{mass}. There is only one light SM-like Higgs $h^0$ from the
real part of $\phi^0_1$. The masses of all the other scalars are set
by $v_R$ which can be very heavy. From the Lagrangian in Eq.
(\ref{VS}) one can easily obtain the interaction terms among the
scalars. Some of the relevant cubic and quartic scalar interaction
vertexes are listed in Table \ref{interaction}.

\begin{table}[htb]
\begin{center}
\begin{tabular}{|c|c||c|c||c|c||c|c|}
\hline Interaction & Vertex & Interaction & Vertex  & Interaction & Vertex & Interaction &  Vertex \\
\hline $S_D S_D S_{\sigma}S_{\sigma}$  & $- i 2\lambda_{D}$ &
$S_DS_Dh^0$ & $ - i \lambda_{1,D}\kappa$ & $S_D S_D S_{\sigma}$ & $
- i 2 \lambda_{D} v_{\sigma}$ & $S_DS_DH_2^0$ &  $- i \lambda_{3,D}
v_R$
\\
$S_D S_D H H^*$  & $ - i \lambda_{1,D}$ &$S_{\sigma}S_{\sigma} h^0$
& $- i \lambda_{1,D}\kappa$ &$HH^*S_{\sigma}$ &  $- i  \lambda_{1,D}
v_{\sigma}$ &$S_{\sigma}S_{\sigma} H_2^0$   &  $- i \lambda_{3,D}
v_R$
\\
$S_DS_Dh^0H_1^0$        & $- i 2\lambda_{2,D}$ &$S_DS_DH_1^0$ & $- i
2\lambda_{2,D}\kappa$ &$h^0H_1^0S_{\sigma}$ & $- i 2\lambda_{2,D}
v_{\sigma}$ &$S_{\sigma}S_{\sigma}S_{\sigma}$ &  $- i 6\lambda_D
v_{\sigma}$
\\
$S_DS_D\Delta\Delta^*$      & $- i \lambda_{3,D}$
&$S_{\sigma}S_{\sigma} H_1^0$   & $- i 2 \lambda_{2,D} \kappa $
&$\Delta\Delta^*S_{\sigma}$ & $- i  \lambda_{3,D} v_{\sigma}$&
$h^0h^0 H^0_2$&  $- i \alpha_1 v_R$
\\
\hline
\end{tabular}
\end{center}
\caption{The cubic and quartic scalar vertexes  among Higgs singlets
and multiplets, where $HH^*$ stands for any states of $\left (
h^0h^0,H_1^0H_1^0,A_1^0A_1^0,H_2^+H_2^-\right )$ and
$\Delta\Delta^*$ stands for any states of $\left
(H_L^0H_L^{0},A_L^0A_L^0,H_L^+H_L^-,H_L^{++}
H_L^{--},H_2^0H_2^0,H_R^{++}H_R^{--}\right )$. }\label{interaction}
\end{table}

\section{Dark matter signal in the 1HBDM} \label{Sec1HBDM}

\begin{figure}[htb]
\begin{center}
\includegraphics[scale=1.2]{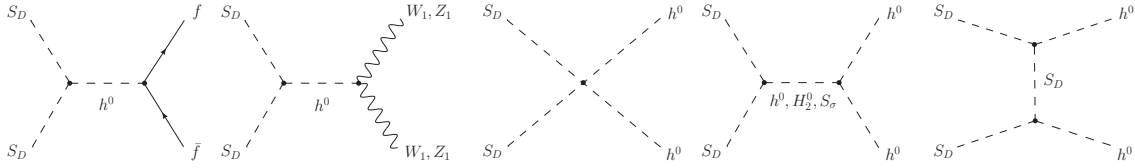}
\end{center}
\caption{Feynman diagrams for the DM annihilation in the 1HBDM.}
\label{Feynman1}
\end{figure}

As discussed in Sec. \ref{Model}, an approximate global $U(1)$
symmetry on $S$ can naturally lead to a light DM mass $m_D$. Here we
focus on $10 \;{\rm GeV} \leq m_D \leq 500$ GeV. Considering the
case $v_\sigma > v_R \sim 10 \; {\rm TeV} \gg \kappa$, one may find
that most of scalar bosons  in Table \ref{mass} are very heavy
except the SM-like one $h^0$. In this case, the possible
annihilation products are $h^0 h^0$, $W_1 W_1 / Z_1 Z_1$ and fermion
pairs $ f \bar f$  as shown in Fig. \ref{Feynman1}. For $s$-channel
annihilation processes, the intermediate particles may be $h^0$,
$H_1^0$, $H_2^0$ and $H_3^0$. Because of $v_L \simeq 0$, one may
neglect the $H_3^0$ case. In addition, the  $H^0_1$ contribution is
also negligible as $m_{H_1^0} \gg m_{h^0}$. For the $f \bar f$
annihilation process, the main contribution comes from the $h^0$
exchange diagram. This is because $H^0_{2}$ dominantly couples to
the very heavy right-handed Majorana neutrinos (the corresponding
annihilation process is kinematically forbidden). For the $W_1 W_1 /
Z_1 Z_1$ processes, the diagram involving $H^0_2$ is suppressed by
$m_{H_2^0} \gg m_{h^0}$. Notice that $S_\sigma$ may be the
intermediate particle for the $h^0 h^0$ case. It is clear that the
dominant annihilation processes in Fig. \ref{Feynman1} are the same
as that in the minimal extension of SM with a real gauge singlet
scalar when $m_D < m_{h^0}$ \cite{McDonald}. In the 1HBDM, the DM
annihilation cross sections $\hat{\sigma}= 4 E_1 E_2 \sigma v$
($E_1$ and $E_2$ are the energies of two incoming DM particles) for
different annihilation channels have the following forms:
\begin{eqnarray}
 \LARGE \hat{\sigma}_{f \bar{f}} &=& \frac{\lambda_{1,D}^2 m_f^2}{4
\pi} \frac{1}{(s-m_{h^0}^2)^2+m_{h^0}^2 \Gamma_{h^0}^2}
\frac{(s-4m_f^2)^{1.5}}{\sqrt{s}}, \label{sigmaff} \\
 \hat{\sigma}_{Z_1 Z_1} &=&  \frac{\lambda_{1,D}^2 }{16 \pi}
\frac{s^2}{(s-m_{h^0}^2)^2+m_{h^0}^2 \Gamma_{h^0}^2} \sqrt{1-\frac{4
m_{Z_1}^2}{s}} \left(1- \frac{4m_{Z_1}^2}{s}+ \frac{12
m_{Z_1}^4}{s^2} \right),\\
 \hat{\sigma}_{W_1 W_1} &=&  \frac{\lambda_{1,D}^2 }{8 \pi}
\frac{s^2}{(s-m_{h^0}^2)^2+m_{h^0}^2 \Gamma_{h^0}^2} \sqrt{1-\frac{4
m_{W_1}^2}{s}} \left(1- \frac{4m_{W_1}^2}{s}+ \frac{12
m_{W_1}^4}{s^2} \right),\\
 \hat{\sigma}_{h^0 h^0} &=&  \frac{\lambda_{1,D}^2 }{16 \pi}
\sqrt{1-\frac{4 m_{h^0}^2}{s}} \left[ G_1^2 - \frac{8 \lambda_{1,D}
\kappa^2}{s-2 m_{h^0}^2} G_1 F(\xi_{h^0}) + \frac{8 \lambda_{1,D}^2
\kappa^4}{(s-2 m_{h^0}^2)^2} \left( \frac{1}{1-\xi_{h^0}^2} +
F(\xi_{h^0})\right) \right], \label{sigmahh}
\end{eqnarray}
where $s$ is the squared center-of-mass energy \cite{modify}. The
quantity $F$ is defined as
$F(\xi_{h^0})\equiv\mbox{arctanh}(\xi_{h^0})/\xi_{h^0}$ with
$\xi_{h^0} = \sqrt{s- 4 m_D^2} \sqrt{s- 4 m_{h^0}^2}/(s- 2
m_{h^0}^2)$. The Higgs decay width $\Gamma_{h^0}$ and $G_1$ are
given by
\begin{eqnarray} \Gamma_{h^0} & =& \frac{\sum m_f^2 }{8 \pi
\kappa^2} \frac{(m_{h^0}^2 - 4 m_f^2)^{1.5}}{m_{h^0}^2} +
\frac{m_{h^0}^3}{16 \pi \kappa^2} \sqrt{1- \frac{4
m_{W_1}^2}{m_{h^0}^2}}\left(1- \frac{4m_{W_1}^2}{m_{h^0}^2}+
\frac{12 m_{W_1}^4}{m_{h^0}^4} \right)
 \nonumber\\ & &   + \frac{m_{h^0}^3}{32 \pi \kappa^2} \sqrt{1- \frac{4
m_{Z_1}^2}{m_{h^0}^2}}\left(1- \frac{4m_{Z_1}^2}{m_{h^0}^2}+
\frac{12 m_{Z_1}^4}{m_{h^0}^4} \right)+ \frac{\lambda_{1,D}^2
\kappa^2}{32 \pi} \frac{\sqrt{m_{h^0}^2 - 4
m_D^2}}{m_{h^0}^2} \;,\nonumber \\
G_1 &=& 1 + \frac{3 m_{h^0}^2}{s - m_{h^0}^2} + \frac{\alpha_1
\lambda_{3,D} v_R^2}{s - m_{H_2^0}^2} \frac{1}{\lambda_{1,D}} +
\frac{m_\sigma^2}{s- m_\sigma^2}\;.  \label{G1}
\end{eqnarray}
From Eqs. (\ref{sigmaff}-\ref{G1}) seven unknown parameters enter
the expression of total annihilation cross section, namely,
$m_{h^0}$, $m_D$, $\lambda_{1,D}$, $\alpha_1\lambda_{3,D}$,
$m_\sigma^2$, $m_{H_2^0}^2$ and $v_R$. For the mass of SM-like
Higgs, we take $m_{h^0} = 120$ GeV in the following parts.  In fact,
one may neglect the squared center-of-mass energy $s$ in the terms
$s - m_{H_2^0}^2$ and $s- m_\sigma^2$ since  the masses of
$s_\sigma$ and $H^0_{2}$ are around $v_R$. In a good approximation,
we find that only three independent parameters
\begin{eqnarray}
m_D, \lambda_{1,D} \; {\rm and} \; \lambda_R \equiv
\alpha_1\lambda_{3,D}/(2\rho_1)
\end{eqnarray}
are relevant to our numerical analysis. Here we have used
$m_{H_2^0}^2 = 2 \rho_1 v_R^2$ as it is shown in Table \ref{mass}.

\subsection{Constraints from the DM relic density} \label{1HBDMA}

In order to obtain the correct DM abundance, one should resolve the
following Boltzmann equation \cite{KOLB}:
\begin{eqnarray}
\frac{d Y}{d x} = - \frac{x \; {\bf s}(x)}{H} \langle \sigma v
\rangle (Y^2 -Y_{EQ}^2) \; , \label{bol}
\end{eqnarray}
where $Y \equiv n/{\bf s}(x)$  denotes the DM number density. The
entropy density ${\bf s}(x)$ and the Hubble parameter $H$ evaluated
at $x=1$ are given by
\begin{eqnarray}
{\bf s}(x) = \frac{2 \pi^2 g_*}{45} \frac{m_D^3}{x^3} \;, \; H =
\sqrt{\frac{4 \pi^3 g_*}{45}} \frac{m_D^2}{M_{\rm PL}} \label{h} \;,
\end{eqnarray}
where $M_{\rm PL} \simeq 1.22 \times 10^{19}$ GeV is the Planck
energy. $g_*$ is the total number of effectively relativistic
degrees of freedom. The numerical results of $g_*$ have been
presented in Ref. \cite{APP}. Here we take the QCD phase transition
temperature to be 150 MeV. The thermal average of the annihilation
cross section times the relative velocity $\langle \sigma v \rangle$
is a key quantity in the determination of the DM cosmic relic
abundance. We adopt the usual single-integral formula for $\langle
\sigma v \rangle$ \cite{Edsjo:1997bg}:
\begin{eqnarray}
\langle \sigma v \rangle = \frac{1}{n_{EQ}^2} \frac{m_D}{64 \pi^4 x}
\int_{4 m_D^2}^{\infty} \hat{\sigma}(s) \sqrt{s} K_1(\frac{x
\sqrt{s}}{m_D}) d s \;, \label{cross}
\end{eqnarray}
with
\begin{eqnarray}
n_{EQ} = \frac{g_i}{2 \pi^2} \frac{m_D^3}{x} K_2(x) \; , \;
\hat{\sigma}(s) = \hat{\sigma} \; g_i^2 \; \sqrt{1-\frac{4
m_D^2}{s}} \; , \label{sigmahat}
\end{eqnarray}
where $K_1(x)$ and $K_2(x)$ are the modified Bessel functions.
$x\equiv m_D/T$ and $g_i =1$ is the internal degrees of freedom for
the scalar dark matter $S_D$. In terms of the annihilation cross
section $\hat{\sigma}$ in Eqs. (\ref{sigmaff}-\ref{sigmahh}), one
can numerically calculate the thermally averaged annihilation cross
section $\langle \sigma v \rangle$. Finally, we may obtain the DM
relic density $\Omega_D h^2 = 2.74 \times 10^8 \; Y_0 \; m_D /{\rm
GeV}$ by use of the result $Y_0$ of the integration of Eq.
(\ref{bol}).

When the DM mass $m_D$ is larger than the mass of top quark, one
will not meet the resonance \cite{BW} and threshold \cite{Threshold}
effects in our model. Thus we use the approximate formulas to
calculate the DM relic density for $ 200 \,{\rm GeV} \leq m_D \leq
500\, {\rm GeV}$. In this case, $\langle \sigma v \rangle$ can be
expanded in powers of relative velocity and $x^{-1}$ for
nonrelativistic gases. To the first order $\langle \sigma v \rangle
\simeq \sigma_0 x^{-n}$, where $n=0(1)$ for $s(p)$-wave annihilation
process \cite{KOLB}. The approximate formula for $\langle\sigma
v\rangle$ is given by \cite{Srednicki:1988ce}
\begin{eqnarray}
\langle \sigma v \rangle = \sigma_0 x^{-n} = \frac{1}{m_D^2} \left [
\omega - \frac{3}{2} (2 \omega - \omega ')x^{-1} + \ldots
\right]_{s/4m_D^2=1} \; , \label{Expand}
\end{eqnarray}
where $\omega = (\hat{\sigma}_{f \bar{f}} + \hat{\sigma}_{Z_1 Z_1} +
\hat{\sigma}_{W_1 W_1} + \hat{\sigma}_{h^0 h^0})/4$ and the prime
denotes derivative with respect to $s/4m_D^2$. $\omega$ and its
derivative are all to be evaluated at $s/4m_D^2=1$. Then $\Omega_{D}
h^2$ is given by \cite{KOLB}
\begin{eqnarray}
\Omega_{D} h^2 = 1.07 \times 10^9 \, \frac{(n+1)
x_f^{n+1}}{g_*^{1/2} M_{\rm PL} \, \sigma_0} \, {\rm GeV}^{-1}
\label{nin}
\end{eqnarray}
with
\begin{eqnarray}
x_f  =  {\rm ln}[0.038(n+1)(g_{i}/g_*^{1/2}) M_{\rm PL} m_D
\sigma_0] - (n+ 1/2)\, {\rm ln} \{ {\rm ln}
[0.038(n+1)(g_{i}/g_*^{1/2}) M_{\rm PL}  m_D  \sigma_0] \}.
\label{xf}
\end{eqnarray}
Notice that we take $g_* = 345/4$ for $ 200 \,{\rm GeV} \leq m_D
\leq 500\, {\rm GeV}$.

\begin{figure}[htb]
\begin{center}
\includegraphics[width=7cm,height=6cm,angle=0]{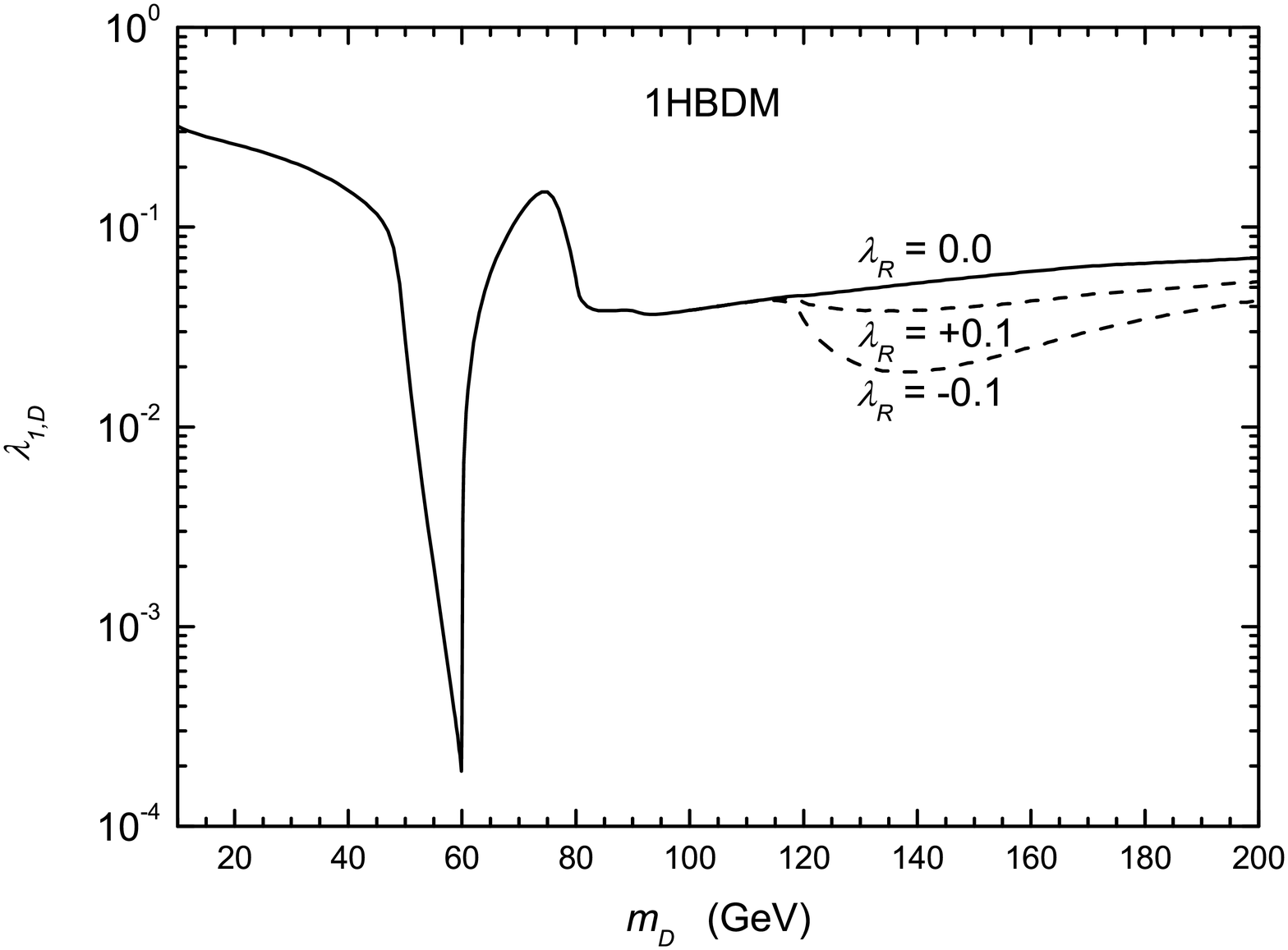}\includegraphics[width=7cm,height=6cm,angle=0]{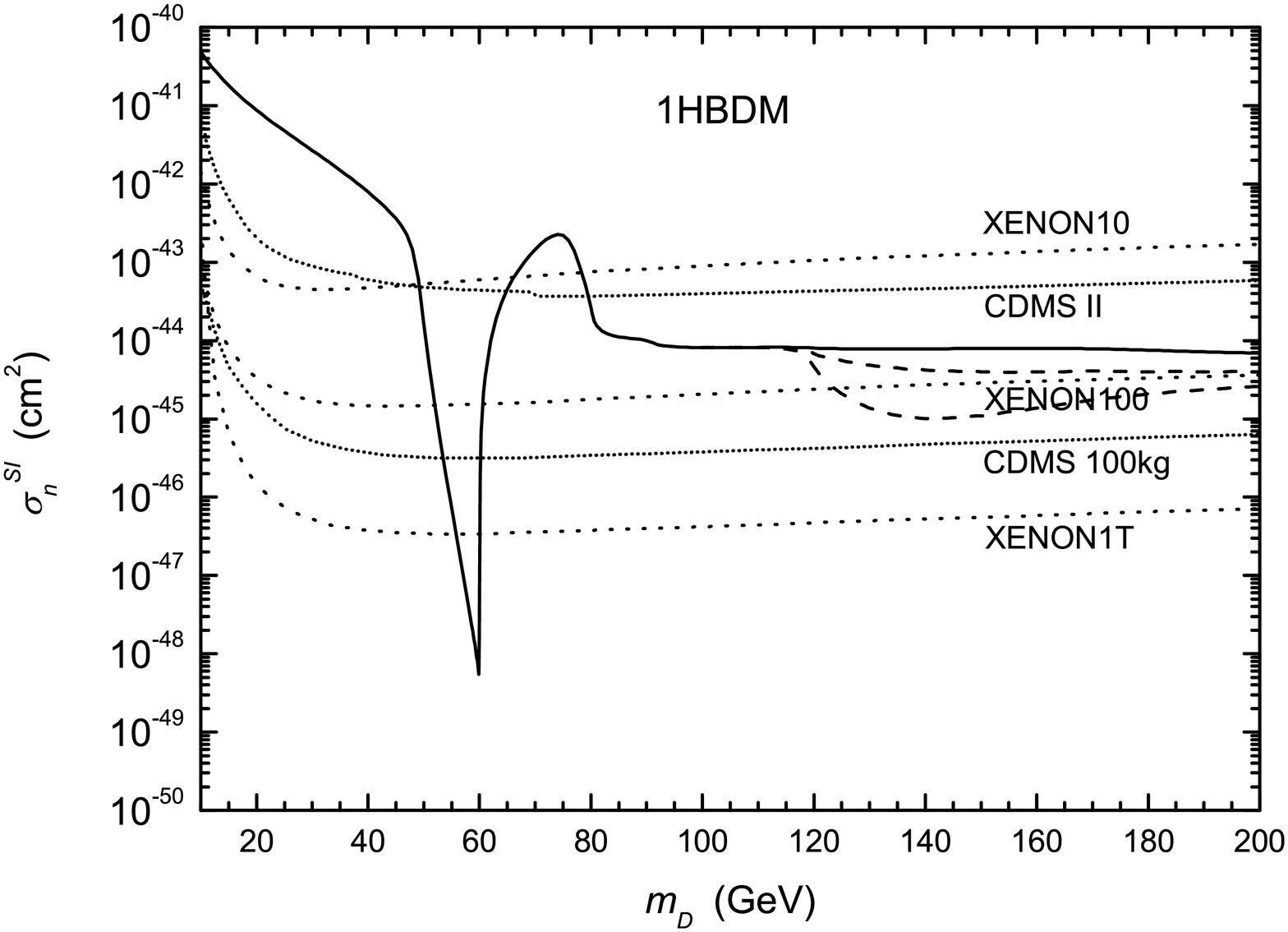}
\includegraphics[width=7cm,height=6cm,angle=0]{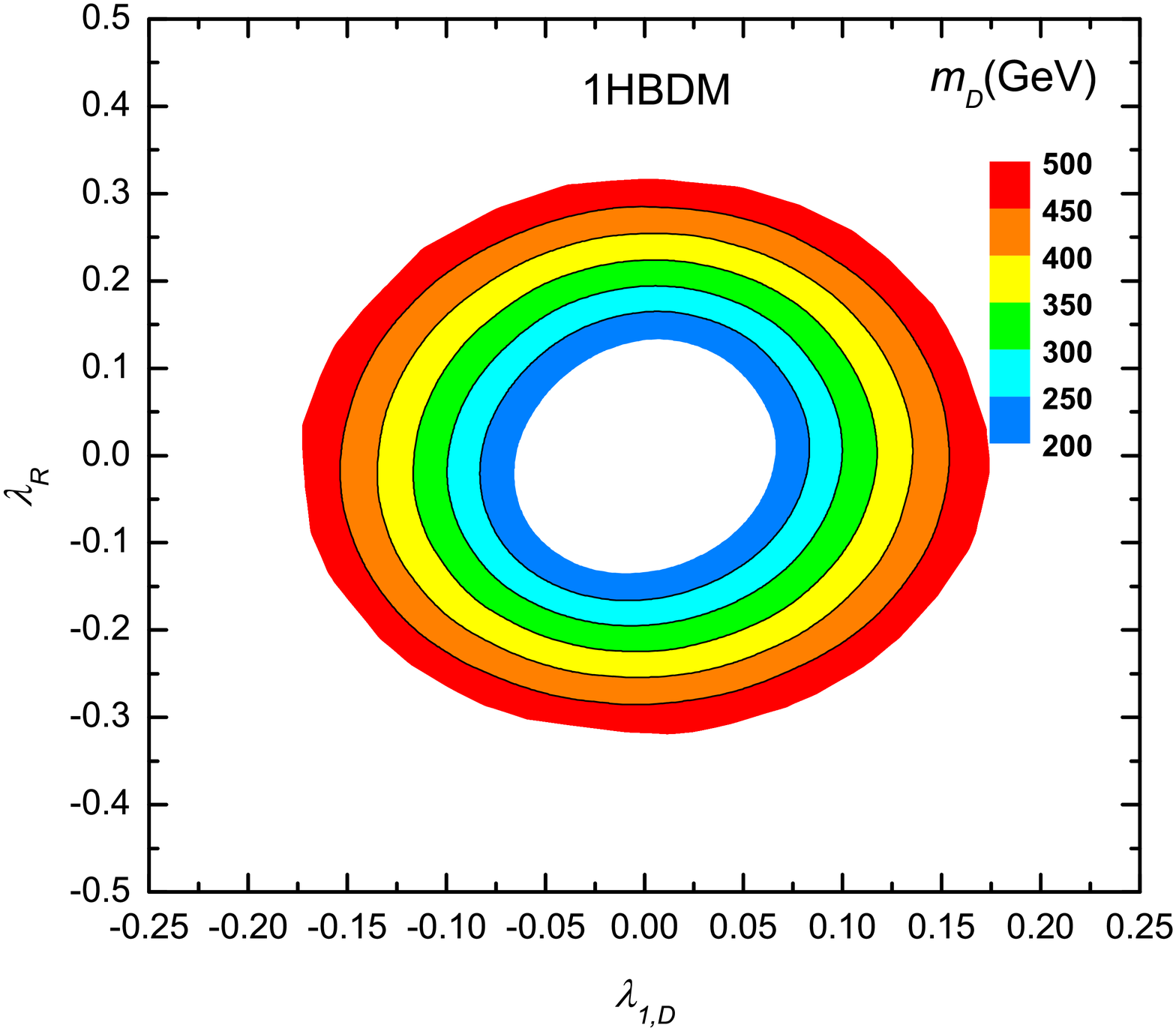}\includegraphics[width=7cm,height=6cm,angle=0]{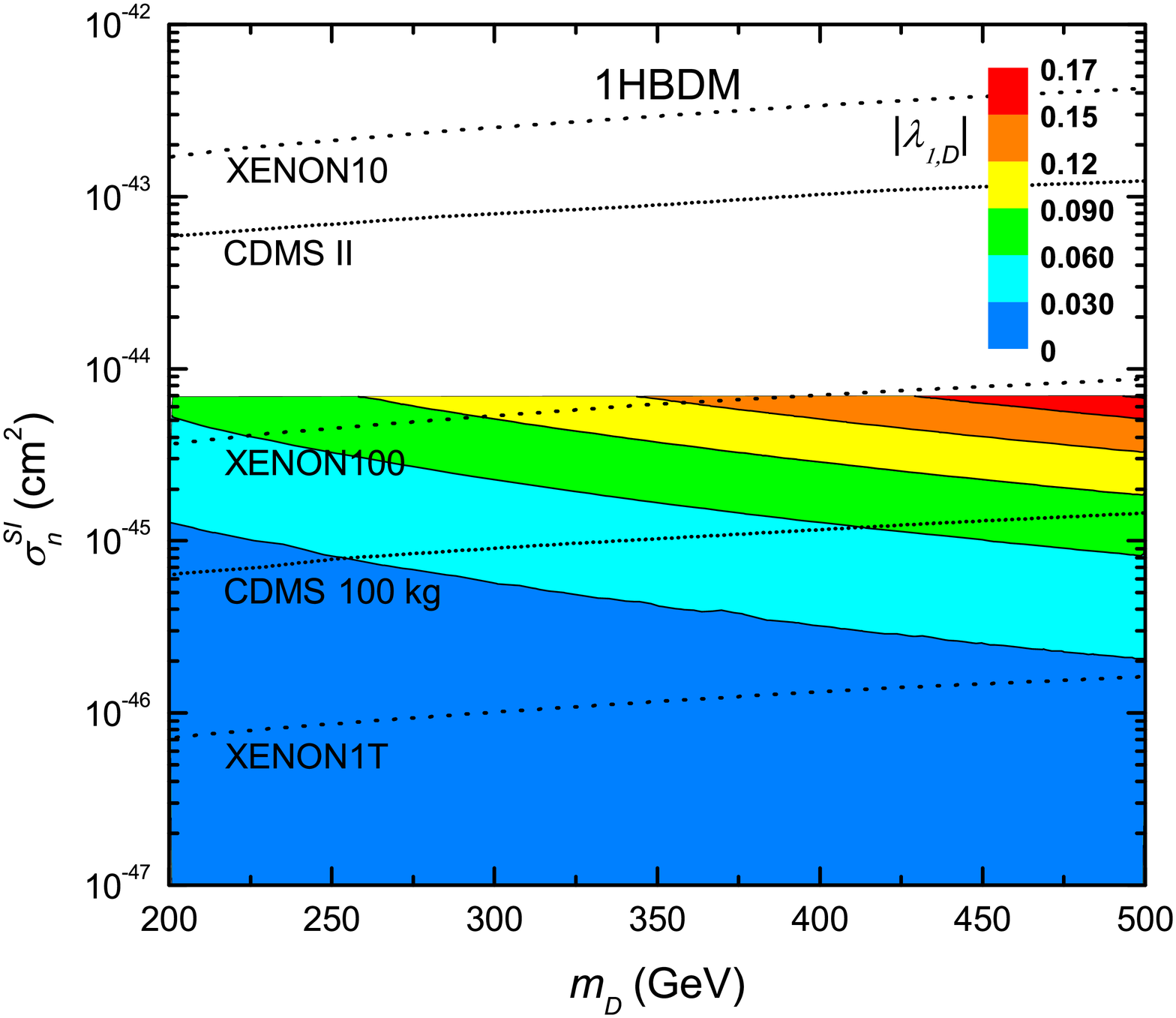}
\end{center}
\caption{Left panels: the predicted coupling $\lambda_{1,D}$ as a
function of $\lambda_R$ and the DM mass $m_D$ from the observed DM
abundance in the 1HBDM. Right panels: the predicted DM-nucleon
scattering cross section $\sigma_{n}^{SI}$ in the 1HBDM with current
and future experimental upper bounds. } \label{1HBDM}
\end{figure}

In terms of the observed DM abundance $0.1088 \leq\Omega_D h^2 \leq
0.1158$ \cite{WMAP7}, we numerically solve the Boltzmann equation
and derive the coupling $\lambda_{1, D}$ with different $\lambda_R$
for $10 \,{\rm GeV} \leq m_D \leq 200\, {\rm GeV}$. The numerical
results are shown in Fig. \ref{1HBDM} (upper-left panel). Due to the
resonance contribution,  a very small value of the coupling
$\lambda_{1, D}$ can be derived from the observed DM abundance for
the resonance region ($0.8 \; m_{h^0} \lesssim 2 m_D < m_{h^0}$).
Except for the resonance region, one may find $\lambda_{1,D} \sim
{\cal O} (10^{-2} - 10^{-1})$. The parameter $\lambda_R$ plays an
important role to determine the DM relic density if $m_D > m_{h^0}$.
For illustration, we also plot the $\lambda_R = \pm 0.1$ cases which
can significantly change the predicted $\lambda_{1, D}$ as shown in
Fig. \ref{1HBDM}. In fact, $\lambda_{1, D}$ may be very small (even
to be zero) for the larger $|\lambda_R|$. In this case, the
$H_2^0$-exchange annihilation process is dominant. Here we have
assumed $\lambda_{1, D}$ is positive. If we simultaneously change
the signs of $\lambda_{1, D}$ and $\lambda_R$, the negative
$\lambda_{1, D}$ case may be approximately induced from the positive
case. This feature can be well understood from Eqs.
(\ref{sigmahh}-\ref{G1}). It should be mentioned that the thermally
averaged annihilation cross section $\langle \sigma v \rangle$ will
significantly change as the evolution of the Universe when the DM
particle is nearly one-half the mass of a resonance \cite{BW}. This
is the Breit-Wigner resonance effect which has been used to explain
the recent PAMELA \cite{PAMELA}, ATIC \cite{ATIC} and Fermi
\cite{Fermi} anomalies. Notice that the decaying $S_D$ with a
lifetime around ${\cal O} (10^{26}s)$ can also account for the
electron and positron anomalies \cite{Guo:2010vy}. Here we have
considered the Breit-Wigner resonance effect for the determination
of the coupling $\lambda_{1, D}$.

For $ 200 \,{\rm GeV} \leq m_D \leq 500\, {\rm GeV}$, we use the
approximate formulas to scan the whole parameter space $\lambda_{1,
D}$ and $\lambda_R$.  The allowed parameter space is shown in Fig.
\ref{1HBDM} (lower-left panel), which gives an allowed range $-0.17
\lesssim \lambda_{1, D} \lesssim 0.17$ and $-0.32 \lesssim \lambda_R
\lesssim 0.32.$ The central region of this figure is excluded since
these points can not provide large enough annihilation cross section
to give the desired DM abundance. Notice that the approximate global
symmetry $U(1)$ requires $m_D^2/v_R^2 \ll \lambda_{1,D}$ which means
the region near $\lambda_{1,D}=0$ is disfavored.

\subsection{Dark matter direct search}

For the scalar dark matter, the DM elastic scattering cross section
on a nucleon is spin-independent, which is given by \cite{DM}
\begin{eqnarray}
\sigma_{n}^{SI}  \approx \frac{4}{\pi} \left (\frac{m_D \; m_n}{m_D+
m_n } \right)^2 \frac{(Z f_p + (A-Z) f_n)^2}{A^2} \; ,
\end{eqnarray}
where $m_n$ is the nucleon mass. $Z$ and $A-Z$ are the numbers of
protons and neutrons in the nucleus. $f_{p,n}$ is the coupling
between DM and  protons or neutrons, given by
\begin{eqnarray}
f_{p,n}= \sum_{q=u,d,s} f_{Tq}^{(p,n)} a_q \frac{m_{p,n}}{m_q} +
\frac{2}{27} f_{TG}^{(p,n)}  \sum_{q=c,b,t} a_q \frac{m_{p,n}}{m_q},
\label{fn}
\end{eqnarray}
where $f_{Tu}^{(p)}=0.020 \pm 0.004$, $f_{Td}^{(p)}=0.026 \pm
0.005$, $f_{Ts}^{(p)}=0.118 \pm 0.062$, $f_{Tu}^{(n)}=0.014 \pm
0.003$, $f_{Td}^{(n)}=0.036 \pm 0.008$ and $f_{Ts}^{(n)}=0.118 \pm
0.062$ \cite{Ellis:2000ds}. The coupling $f_{TG}^{(p,n)}$ between DM
and gluons from heavy quark loops is obtained from $f_{TG}^{(p,n)} =
1 - \sum_{q=u,d,s} f_{Tq}^{(p,n)}$, which leads to $f_{TG}^{(p)}
\approx 0.84 $ and $f_{TG}^{(n)} \approx 0.83$. In our model, the
DM-quark coupling $a_q$ in Eq. (\ref{fn}) is given by
\begin{eqnarray}
a_q = \frac{\lambda_{1,D} \; m_q}{2 m_D \; m_{h^0}^2}  \,.
\label{aq}
\end{eqnarray}
Because of  $f_n \approx f_p$, we can derive
\begin{eqnarray}
\sigma_{n}^{SI}  \approx \frac{4}{\pi} \left (\frac{m_D \; m_n}{m_D+
m_n } \right)^2 f_n^2 \; .
\end{eqnarray}
It is worthwhile to stress that $\sigma_{n}^{SI}$ is independent of
$\lambda_R$.

Using the predicted $\lambda_{1,D}$ from the observed DM abundance,
we straightly calculate the spin-independent DM-nucleon elastic
scattering cross section $\sigma_{n}^{SI}$. The numerical results
are shown in Fig. \ref{1HBDM} (right panels). For $10 \,{\rm GeV}
\leq m_D \leq 200\, {\rm GeV}$, we find that two DM mass ranges can
be excluded by the current DM direct detection experiments CDMS II
\cite{CDMSII} and XENON10 \cite{XENON10}. Due to the existence of
$\lambda_R$, we can obtain different values of $\sigma_{n}^{SI}$ for
a given DM mass $m_D$ when the annihilation channel $S_D S_D
\rightarrow h^0 h^0$ is open. In this case, one can obtain
$\sigma_{n}^{SI} \lesssim 7\times 10^{-45} {\rm cm}^2$ for $200
\,{\rm GeV} \leq m_D \leq 500\, {\rm GeV}$  as shown in Fig.
\ref{1HBDM} (lower-right panel), which is below the current
experimental upper bounds. Nevertheless the future experiments
XENON100 \cite{XENON100P}, CDMS 100 kg \cite{CDMS100} and XENON1T
\cite{XENON1T} can cover most parts of the allowed parameter space.
For the region near the resonance point, the predicted
$\sigma_{n}^{SI}$ is far below the current and future experimental
upper bounds.

\subsection{Dark matter indirect search}

\begin{figure}[htb]
\begin{center}
\includegraphics[width=7cm,height=6cm,angle=0]{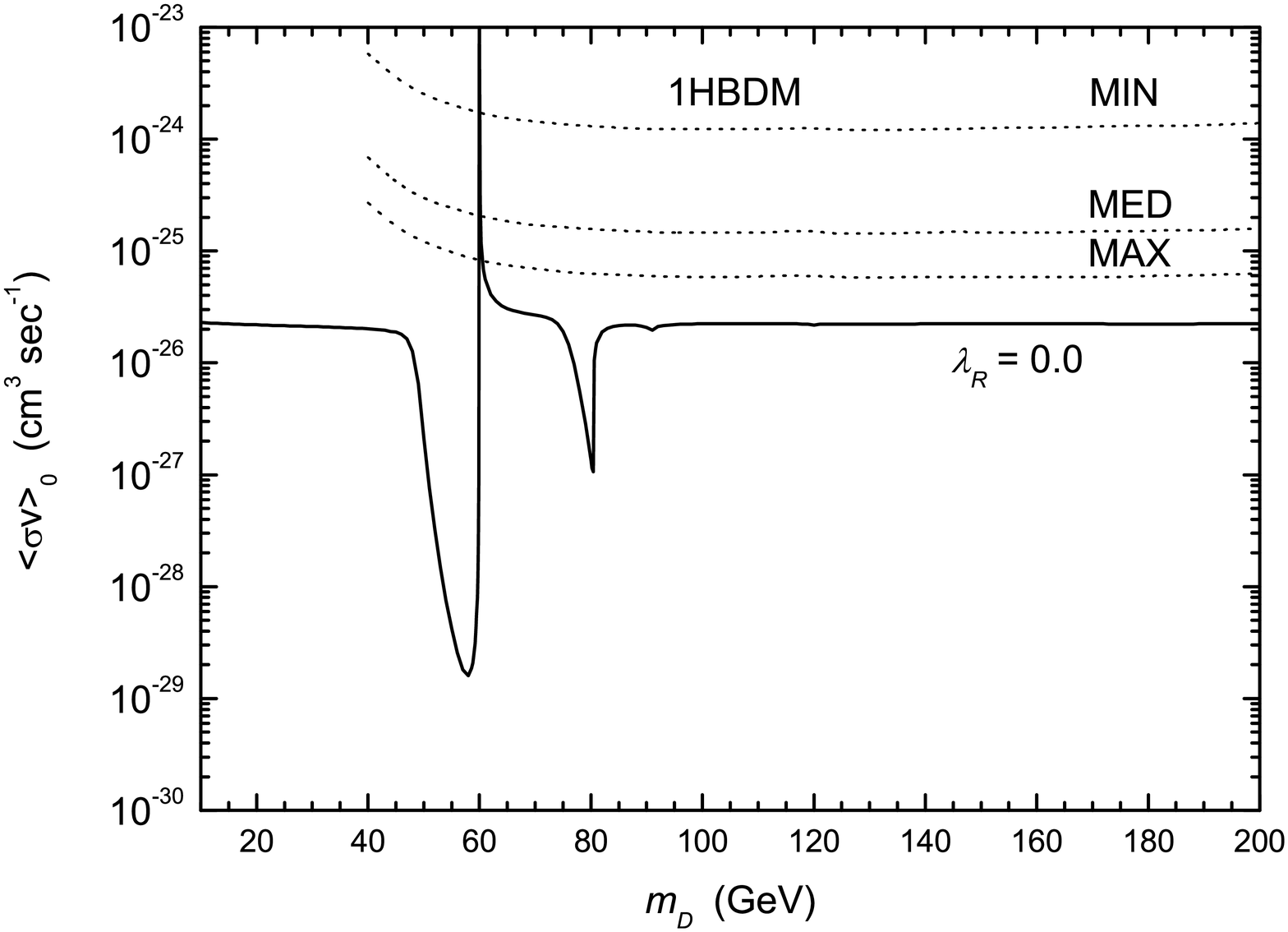}
\end{center}
\caption{  The predicted thermally averaged DM annihilation cross
section $\langle \sigma v \rangle_0$ in the 1HBDM. }
\label{Indirect1}
\end{figure}

As shown in Sec. \ref{1HBDMA}, $\langle \sigma v \rangle$ is a key
quantity in the determination of the DM cosmic relic abundance. On
the other hand, $\langle \sigma v \rangle$ also determines the DM
annihilation rate in the galactic halo. It should be mentioned that
the DM annihilation in the galactic halo occurs at $v \approx
10^{-3}$  ($x \approx 3/v^2 = 3 \times 10^6$). Thus we calculate the
thermally averaged annihilation cross section at $x \approx 3 \times
10^6$, namely $\langle \sigma v \rangle_0$. The numerical results
have been shown in Fig. \ref{Indirect1}  for $ 10 \,{\rm GeV} \leq
m_D \leq 200\, {\rm GeV}$. Notice that we can derive the similar
results for different values of $\lambda_R$. One may find $1 \times
10^{-26} \;{\rm cm}^3 \; {\rm sec}^{-1} \leq\langle \sigma v
\rangle_0 \leq 3 \times 10^{-26} \;{\rm cm}^3 \; {\rm sec}^{-1}$ for
most parts of the parameter space.  The enhanced and suppressed
$\langle \sigma v \rangle_0$ on the two sides of the resonance point
originate from the Breit-Wigner resonance effect \cite{BW}.  When
$m_D$ is slightly less than the $W_1$ boson mass, the channel $S_D
S_D \rightarrow W_1^+ W_1^-$ is open at high temperature, which
dominates the total thermally averaged annihilation cross section
and determines the DM relic density. However this channel is
forbidden in the galactic halo. Thus the threshold effect leads to a
dip around $W_1$ threshold \cite{Threshold}. When $ 200 \,{\rm GeV}
\leq m_D \leq 500\, {\rm GeV}$, one can obtain $\langle \sigma v
\rangle_0 \approx 2.3 \times 10^{-26} \;{\rm cm}^3  \; {\rm
sec}^{-1}$ which is consistent with the usual $s$-wave annihilation
cross section $\langle \sigma v \rangle \approx 3 \times 10^{-26}
\;{\rm cm}^3  \; {\rm sec}^{-1}$ at the freeze-out temperature $x_f
\approx 20$.

In our model, the DM annihilation can generate primary antiprotons
which can be detected by the DM indirect search experiments.
Recently, the PAMELA collaboration reports that the observed
antiproton data is consistent with the usual estimation value of the
secondary antiproton \cite{PAMELA}. Therefore one can use the PAMELA
antiproton measurements to constrain $\langle \sigma v \rangle_0$.
In Fig. \ref{Indirect1}, we have also shown the maximum allowed
$\langle \sigma v \rangle_0$ for the MIN, MED and MAX antiproton
propagation models given in Ref. \cite{Goudelis:2009zz}.   Then we
can find that a very narrow region can be excluded by the PAMELA
antiproton data in our model. In fact, the width of this excluded
region is about $0.4$ GeV for the MED and MAX cases. When double DM
mass $2 m_D$ is slightly less than the Higgs mass $m_{h^0}$, the
predicted $\sigma_{n}^{SI}$ and $\langle \sigma v \rangle_0$ are
very small which means that it is very difficult to detect the DM
signals.

\section{Dark matter signal in the 2HBDM} \label{Sec2HBDM}

We have discussed the Higgs singlet $S_D$ as the cold DM candidate
in the 1HBDM. In this section, we generalize the previous
discussions to the 2HBDM in which the other bidoublet $\chi$ mixes
significantly with $\phi$ and $\Delta_{L,R}$.  In this case the SCPV
can be easily realized \cite{Wu:2007kt}. Comparing with the previous
case, the main differences are that there could be more scalar
particles entering the DM annihilation and scattering processes.
Furthermore, the new contributions from these particles may modify
the correlation between the DM annihilation and DM-nucleon elastic
scattering cross sections, which leads to significantly different
predictions from the other singlet scalar DM models and the previous
discussions.

As shown in Eq. (\ref{Higgscomponent}), the second Higgs bidoublet
$\chi$ contains two neutral Higgs contents $\chi_{1,2}^0$. After the
SSB, $\chi_{1,2}^0$ may obtain the VEVs $w_{1,2}/\sqrt{2}$. The
squared sum of all the VEVs including $\kappa_{1,2}$ should still
lead to $v_{\rm EW} =\sqrt{|\kappa_1|^2 + |\kappa_1|^2 +|w_1|^2+
|w_2|^2} \approx 246$ GeV. In general, the 2HBDM includes three
light neutral Higgs bosons  and a pair of charged light Higgs
particles, whose masses are order of the electroweak energy scale.
For simplicity, we consider $\kappa_2 \sim w_2 \sim 0$. In this
case, it is convenient for us to rotate Higgs bidoublets $\phi$ and
$\chi$ into
\begin{eqnarray}
\phi'  = \left ( \begin{matrix} \frac{h_1 + v_{\rm EW}}{\sqrt{2}} &
{\phi'}_2^+ \cr 0 & {\phi'}_2^0 \cr
\end{matrix} \right ) , \; \chi' = \left (
\begin{matrix} \frac{h_2 + i h_3}{\sqrt{2}} & {\chi'}_2^+ \cr H^- & {\chi'}_2^0 \cr
\end{matrix} \right ) \; ,\label{Basis}
\end{eqnarray}
where $H^{\pm}$ are a pair of light charged Higgs bosons. Then one
can diagonalize the mass matrix of three light neutral Higgs
$h_{1,2,3}$ and derive three light neutral Higgs mass eigenstates.
The relation between $h_{1,2,3}$ and three mass eigenstates can be
written as
\begin{equation}
\left ( \begin{matrix} h_1 \cr h_2 \cr h_3 \cr
\end{matrix} \right )  \; =
\; \left (
\begin{matrix} c^{~}_x c^{~}_z & s^{~}_x c^{~}_z & s^{~}_z \cr -
c^{~}_x s^{~}_y s^{~}_z - s^{~}_x c^{~}_y  & - s^{~}_x s^{~}_y
s^{~}_z + c^{~}_x c^{~}_y  & s^{~}_y c^{~}_z \cr - c^{~}_x c^{~}_y
s^{~}_z + s^{~}_x s^{~}_y  & - s^{~}_x c^{~}_y s^{~}_z - c^{~}_x
s^{~}_y  & c^{~}_y c^{~}_z \cr
\end{matrix} \right ) \; \left ( \begin{matrix} h \cr H \cr A \cr \end{matrix}
\right) \;,
\end{equation}
where $s_x \equiv \sin \theta_x$, $c_x \equiv \cos \theta_x$ and so
on. Due to many unknown parameters in the Higgs potential of 2HBDM,
we can not explicitly calculate three mixing angles $\theta_x,
\theta_y$ and $\theta_z$. For illustration, we consider three
representative cases: (I) $\theta^{~}_x = 60^\circ$, $\theta^{~}_y =
60^\circ$ and $\theta^{~}_z = 150^\circ$; (II) $\theta^{~}_x =
30^\circ$, $\theta^{~}_y = 0^\circ$ and $\theta^{~}_z = 0^\circ$;
(III) $\theta^{~}_x = 0^\circ$, $\theta^{~}_y = 90^\circ$ and
$\theta^{~}_z = 75^\circ$. The Case I means that there is the
significant mixing among three light neutral Higgs. If all $CP$
violation phases are absent, we can obtain $\theta^{~}_y = 0^\circ$
and $\theta^{~}_z = 0^\circ$. In the Case II, the light Higgs $A$ is
$CP$ odd which does not mix with $h$ and $H$. For the Case III, we
only consider the scalar and pseudoscalar mixing, namely
$\theta^{~}_x  = 0^\circ$.

\begin{figure}[htb]
\begin{center}
\includegraphics[scale=1.2]{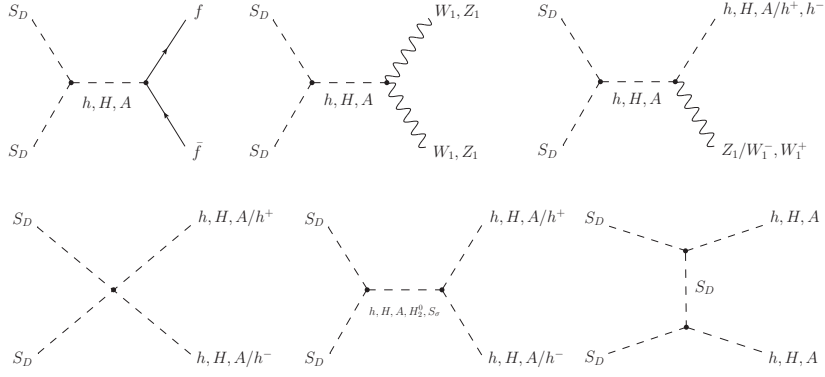}
\end{center}
\caption{ Feynman diagrams for the DM annihilation in the 2HBDM. }
\label{Feynman2}
\end{figure}

In the 2HBDM, the possible DM annihilation products are $f \bar f$,
$W_1 W_1 / Z_1 Z_1$, $W_1^\pm H^\mp/ Z_1 (h, H, A)$, $H^+ H^-$ and
any two of the three neutral states $(h,H,A)$ as shown in Fig.
\ref{Feynman2}.  For a concrete numerical illustration, we choose
all the masses $m_{H}$, $m_{A}$, $m_{H^\pm}= 180$ GeV and
$m_{h}=120$ GeV. For cubic and quartic scalar vertexes, we assume
they are the same as that in the 1HBDM. Namely, the vertexes of $S_D
S_D (h, H, A)$ and $S_D S_D (h, H, A/H^+) (h, H, A/H^-)$ are set
equal to $- i \lambda_{1, D} v_{\rm EW}$ and $- i \lambda_{1, D}$,
respectively. Similarly, the cubic scalar vertexes among the light
Higgs particles $h$, $H, A$ and $H^{\pm}$ are set equal to $- i 3
m_{h}^2/v_{\rm EW}$, and the cubic scalar vertexes between
$S_\sigma$ and two light Higgs particles are assumed to be $- i
\lambda_{1,D} v_\sigma$. It is worthwhile to stress that the heavy
Higgs particles from $\chi'$ may be as the intermediate particles
when two DM candidates annihilate into two light Higgs bosons.
Nevertheless we still can use a coupling $\lambda_R$ to describe the
contributions of all possible heavy Higgs bosons. All annihilation
cross sections $\hat{\sigma}$ have been presented in Appendix
\ref{Appendix.A}.

In the basis of Eq. (\ref{Basis}), the Yukawa interactions for
quarks are given by
\begin{eqnarray}
-{\cal L}_Y  =   \overline{Q_L} \left ( Y^{\phi} \phi'
+\tilde{Y}^{\phi} \tilde{\phi'} + Y^{\chi} \chi' +\tilde{Y}^{\chi}
\tilde{\chi'} \right) Q_R + h.c. ,
\end{eqnarray}
where $Q_{L,R} = (u_{L,R}, d_{L,R})^T$.  When both $P$ and $CP$ are
required to be broken down spontaneously, the Yukawa coupling
matrices $Y^{\phi}$, $\tilde{Y}^{\phi}$, $Y^{\chi}$ and
$\tilde{Y}^{\chi}$ are complex symmetric. Then one may rotate the
quark fields and derive the following Yukawa interactions relevant
to light neutral Higgs particles:
\begin{eqnarray}
-{\cal L}_{LH}  =  \frac{h_1 + v_{\rm EW}}{\sqrt{2}} \left(
\overline{u'_L} Y^{\phi'} u'_R + \overline{d'_L} \tilde{Y}^{\phi'}
d'_R \right) + \frac{h_2 + i h_3}{\sqrt{2}} \overline{u'_L}
Y^{\chi'} u'_R + \frac{h_2 - i h_3}{\sqrt{2}} \overline{d'_L}
\tilde{Y}^{\chi'} d'_R + h.c.,
\end{eqnarray}
where $Y^{\phi'}$ and $\tilde{Y}^{\phi'}$ are diagonal matrixes.
According to the up and down quark masses, we can obtain
$Y^{\phi'}_{qq} = \sqrt{2} m_q / v_{\rm EW}$ and
$\tilde{Y}^{\phi'}_{qq}= \sqrt{2} m_q / v_{\rm EW}$, respectively.
In order to avoid the FCNC processes, we assume $Y^{\chi'}$ and
$\tilde{Y}^{\chi'}$ are approximate diagonal matrixes due to
approximate $U(1)$ family symmetries \cite{Wu} and require
\begin{eqnarray}
Y^{\chi'}_{qq} = R_q \, Y^{\phi'}_{qq} \; {\rm  and} \;
\tilde{Y}^{\chi'}_{qq} = R_q \, \tilde{Y}^{\phi'}_{qq}\;. \label{R}
\end{eqnarray}
Since $Y^{\chi'}$ and $\tilde{Y}^{\chi'}$ don't contribute the quark
masses,  the parameter $R_q$ may be very large except the top quark
case.

\begin{figure}[htb]
\begin{center}
\includegraphics[width=7cm,height=6cm,angle=0]{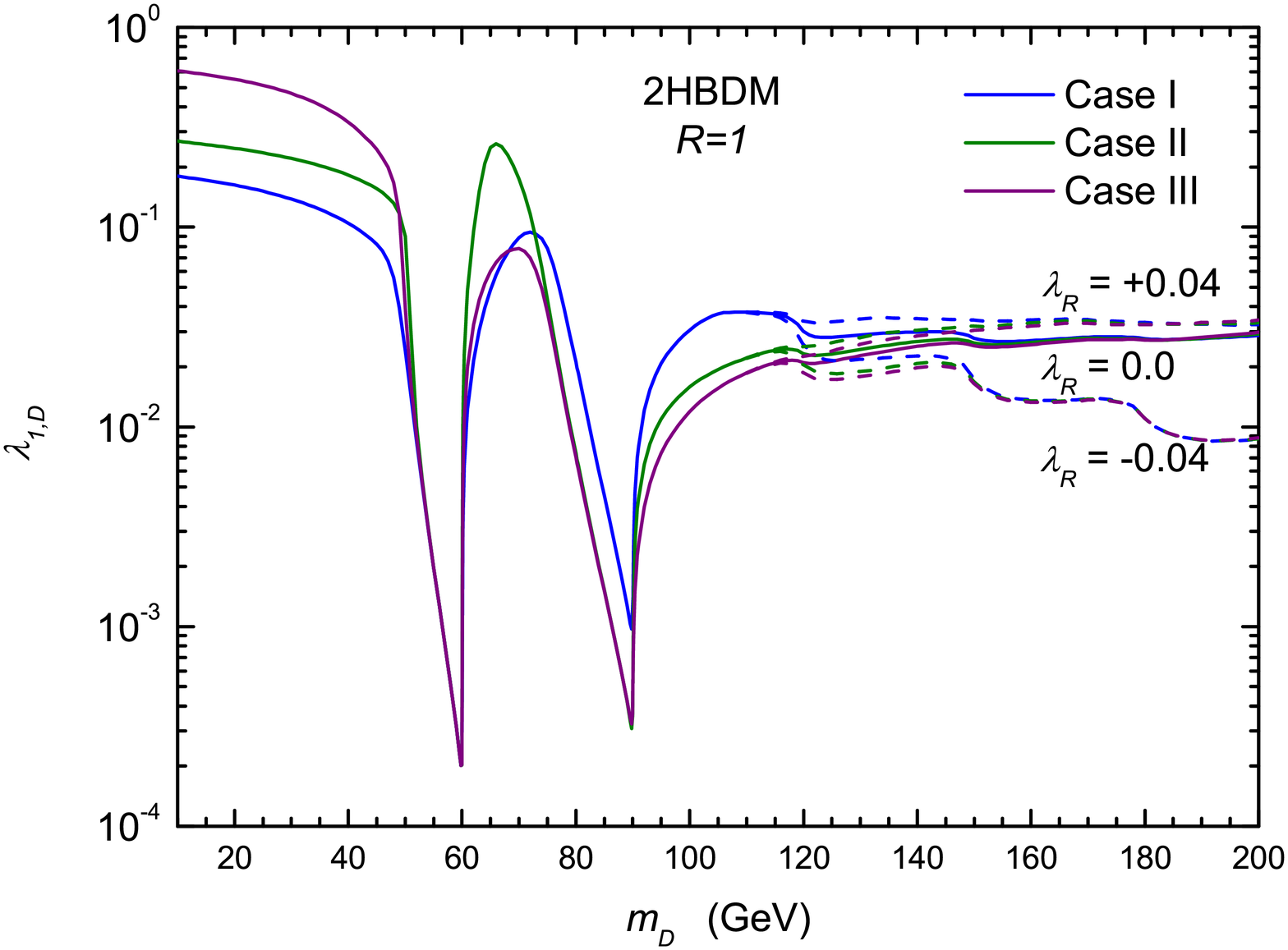}\includegraphics[width=7cm,height=6cm,angle=0]{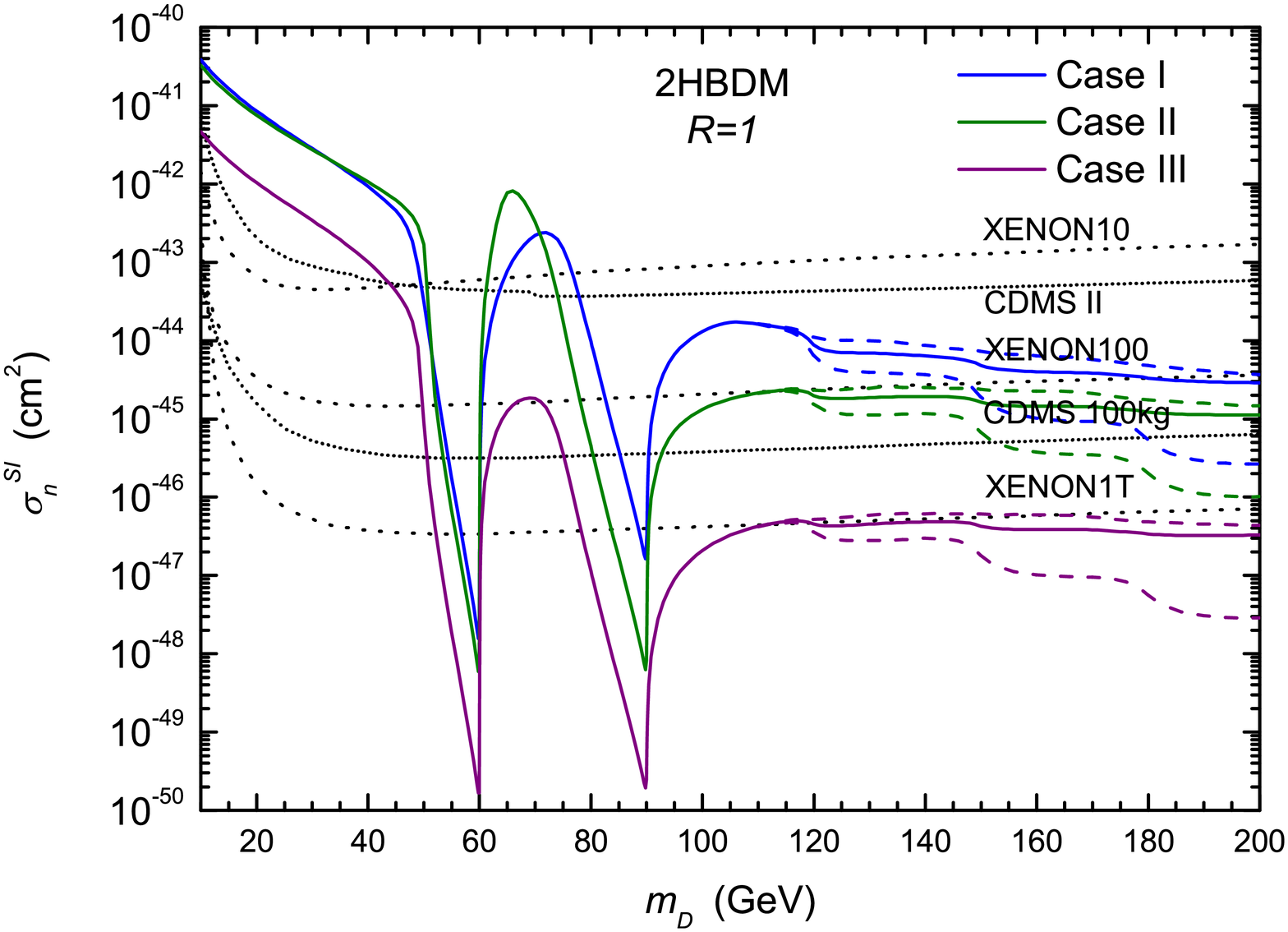}
\includegraphics[width=7cm,height=6cm,angle=0]{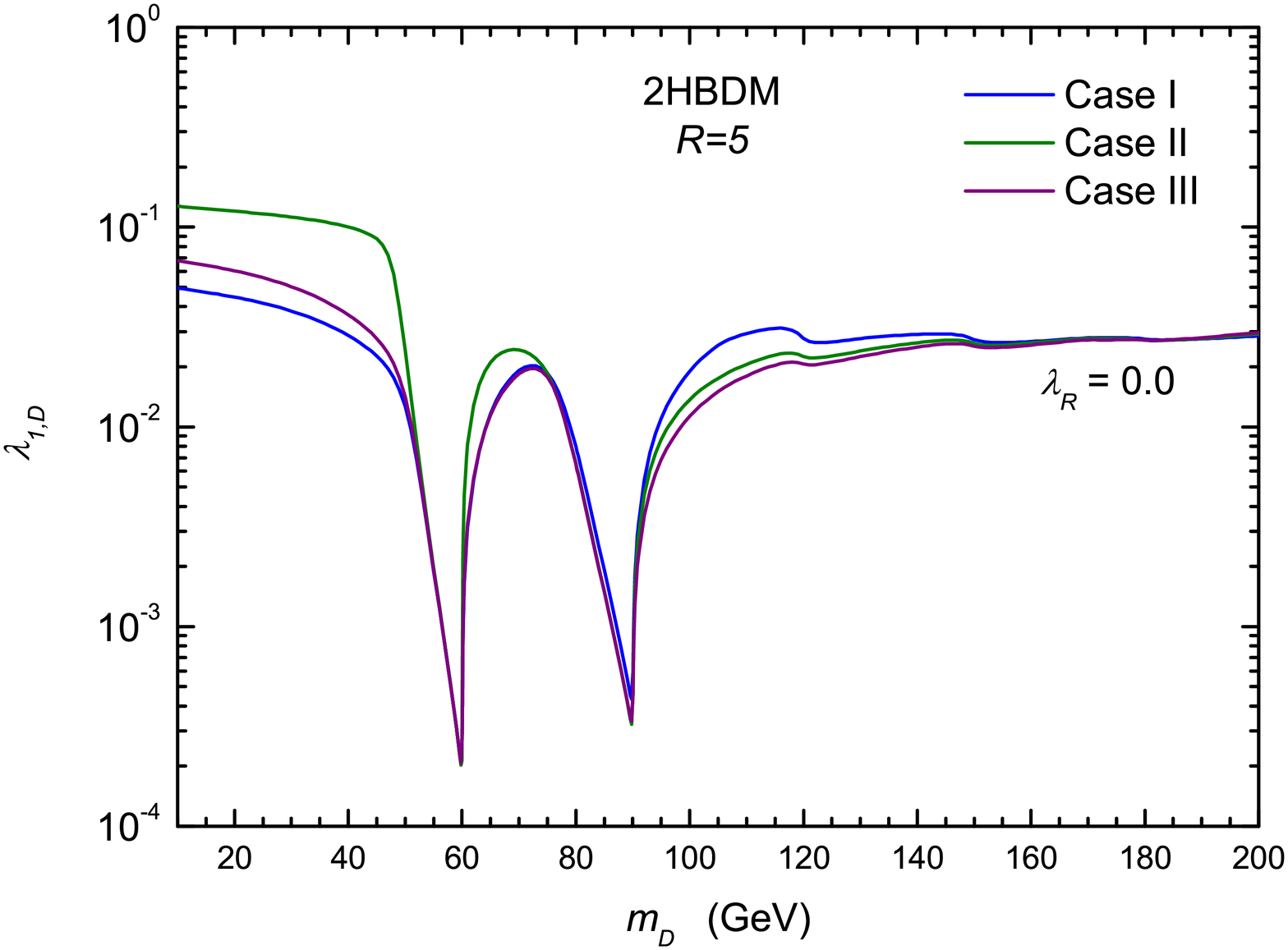}\includegraphics[width=7cm,height=6cm,angle=0]{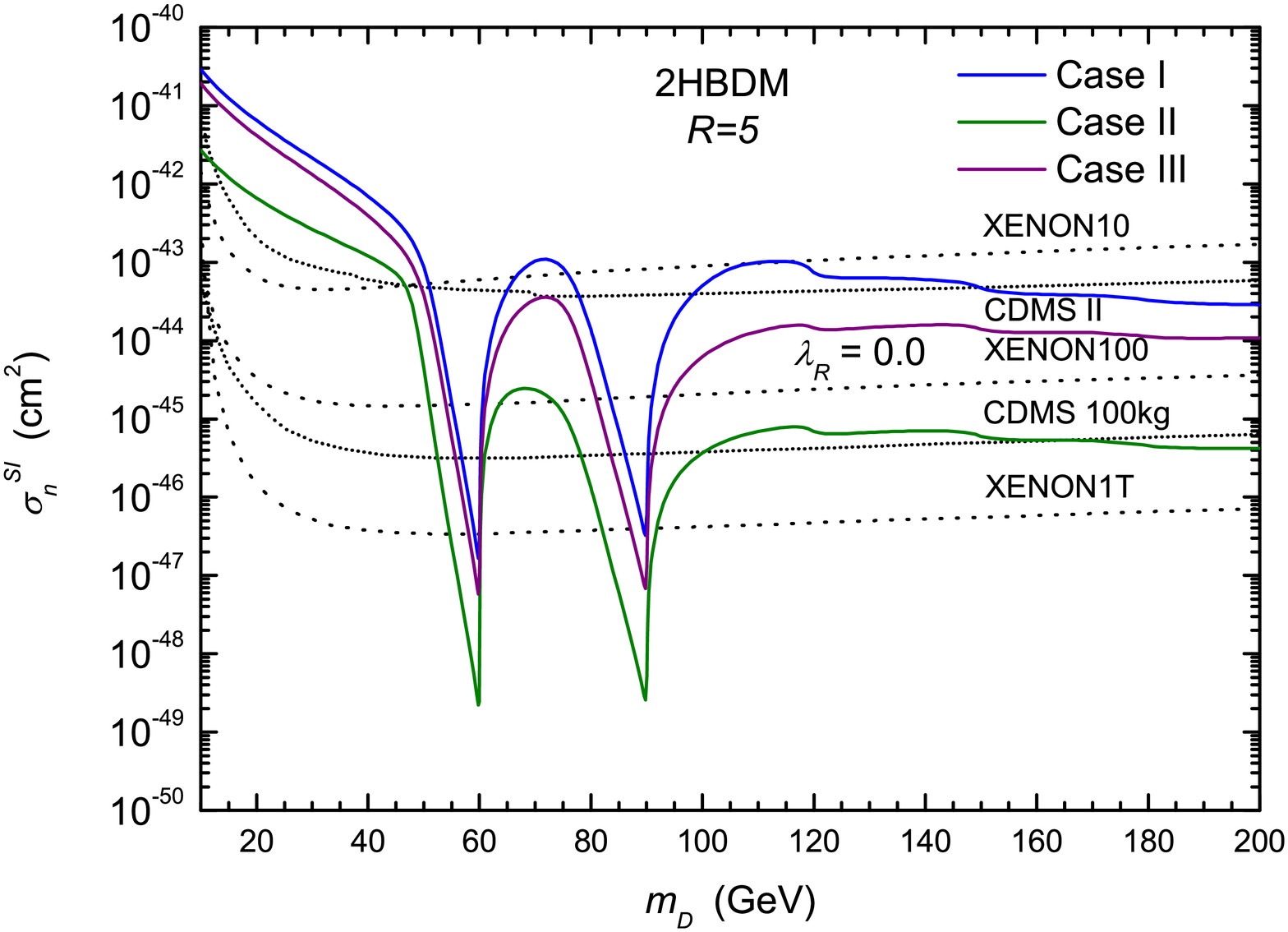}
\end{center}
\caption{ The predicted coupling $\lambda_{1,D}$ and DM-nucleon
scattering cross section $\sigma_{n}^{SI}$ for three mixing cases in
the 2HBDM with $R=1$ and  $R=5$.  } \label{L2HBDM}
\end{figure}

\begin{figure}[t]\begin{center}
\includegraphics[scale=1.0]{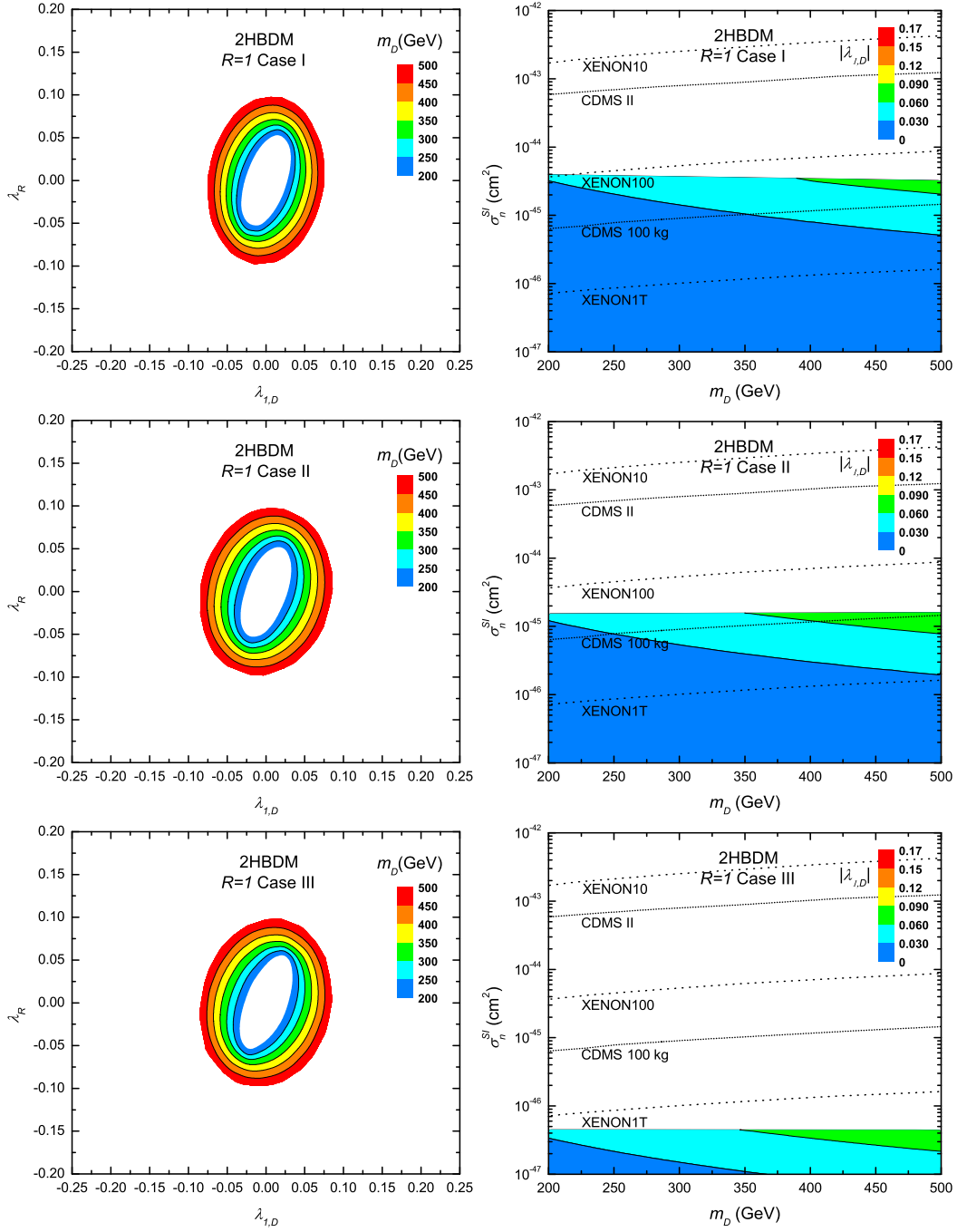}
\end{center}
\vspace{-0.8cm} \caption{The allowed parameter space and the
predicted  $\sigma_{n}^{SI}$ for three mixing cases in the 2HBDM
with $R=1$.} \label{R1}
\end{figure}

\begin{figure}[t]\begin{center}
\includegraphics[scale=1.0]{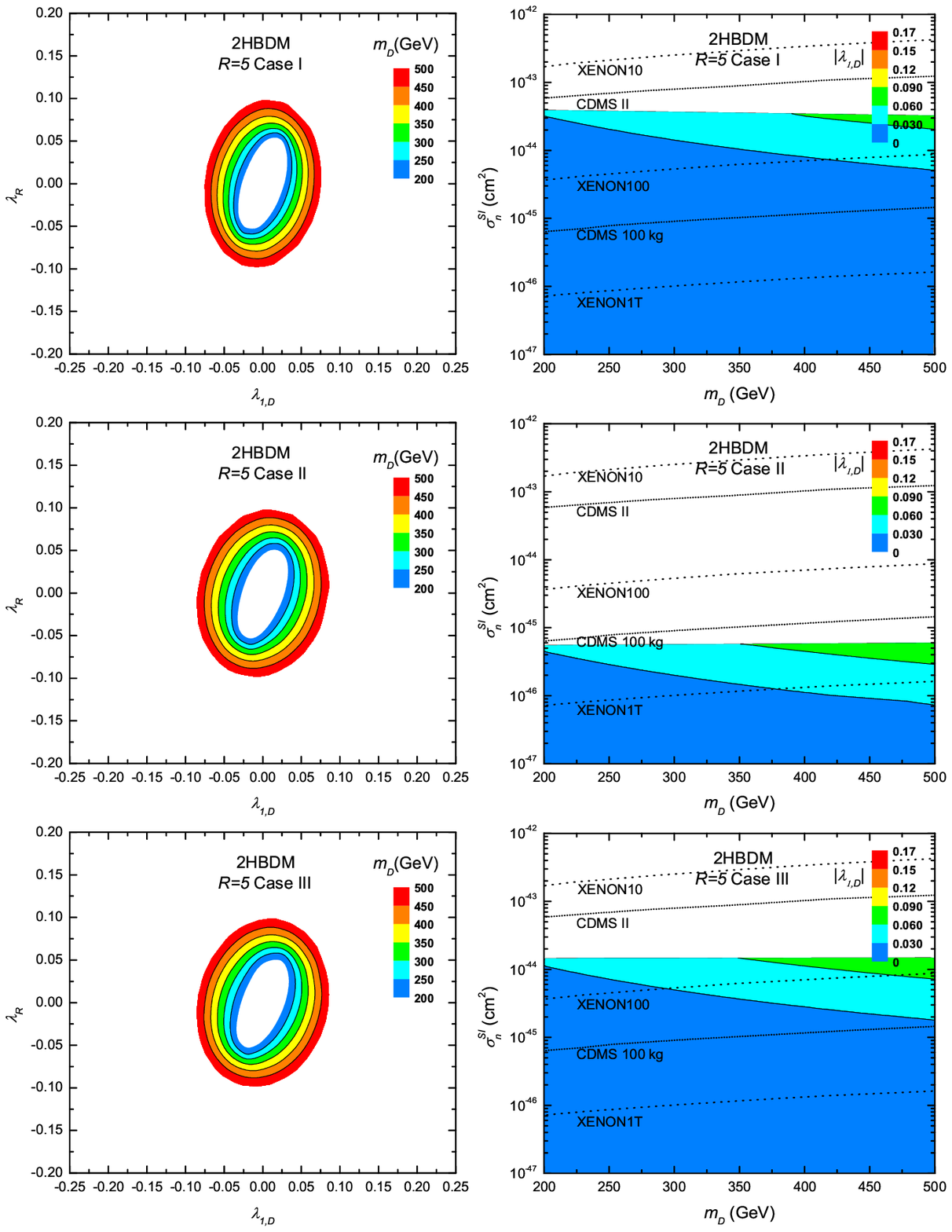}
\end{center}
\vspace{-0.8cm} \caption{The allowed parameter space and the
predicted  $\sigma_{n}^{SI}$ for three mixing cases in the 2HBDM
with $R=5$.} \label{R5}
\end{figure}

In the 2HBDM, the parameter $R_q$ in Eq. (\ref{R}) controls the
Yukawa couplings $Y^{\chi'}_{qq}$ and $\tilde{Y}^{\chi'}_{qq}$.
Furthermore, the parameter $R_q$ will affect the total annihilation
cross section and change the predicted coupling $\lambda_{1,D}$. For
illustration, we choose the following two scenarios
\begin{eqnarray}
R_q \equiv R = 1 \;{\rm and} \; R_q \equiv R =5 \; (q \neq t \;{\rm
and}\; R_t=1\; {\rm for \; the \; top \; quark}) \label{R2}
\end{eqnarray}
to calculate the allowed coupling $\lambda_{1,D}$ from the observed
DM abundance. Considering three kinds of mixing cases and two  $R$
scenarios, we plot the allowed coupling $\lambda_{1,D}$ for $10
\,{\rm GeV} \leq m_D \leq 200\, {\rm GeV}$ in Fig. \ref{L2HBDM}
(left panels). It is clear that $\lambda_{1,D}$ is dependent on the
light Higgs mixing and the parameter $R$ if $m_D < 120$ GeV. When DM
candidate can annihilate into two light Higgs bosons ($m_D \gtrsim
120$ GeV), one can derive the almost same $\lambda_{1,D}$ for three
kinds of mixing cases and two $R$ scenarios, which means that the
light Higgs mixing and the parameter $R$ do not significantly affect
the total annihilation cross section. This conclusion can also be
applied to $200 \,{\rm GeV} \leq m_D \leq 500\, {\rm GeV}$ as shown
in Figs. \ref{R1} and \ref{R5} (left panels).

\begin{figure}[htb]
\begin{center}
\includegraphics[width=7cm,height=6cm,angle=0]{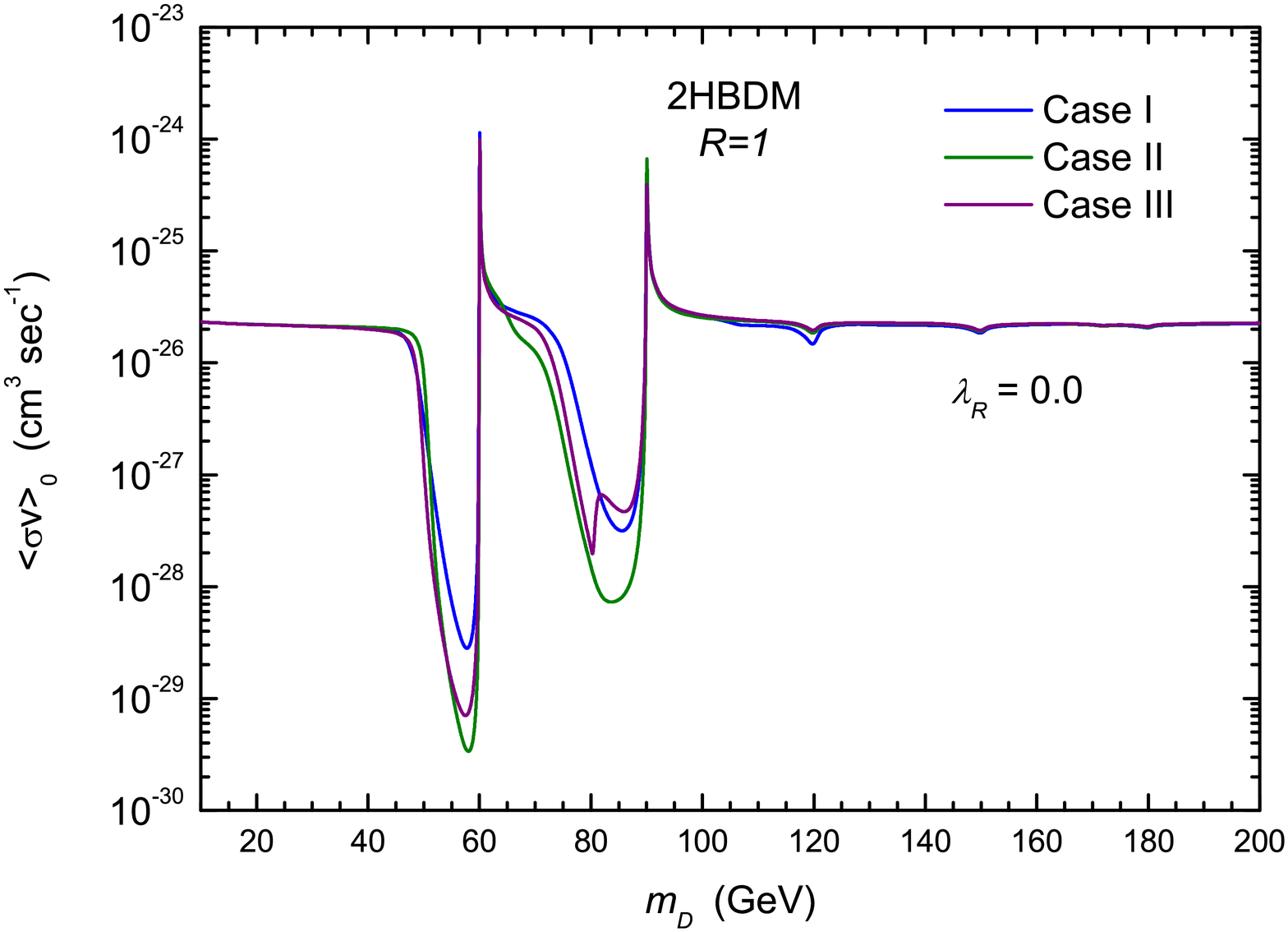}\includegraphics[width=7cm,height=6cm,angle=0]{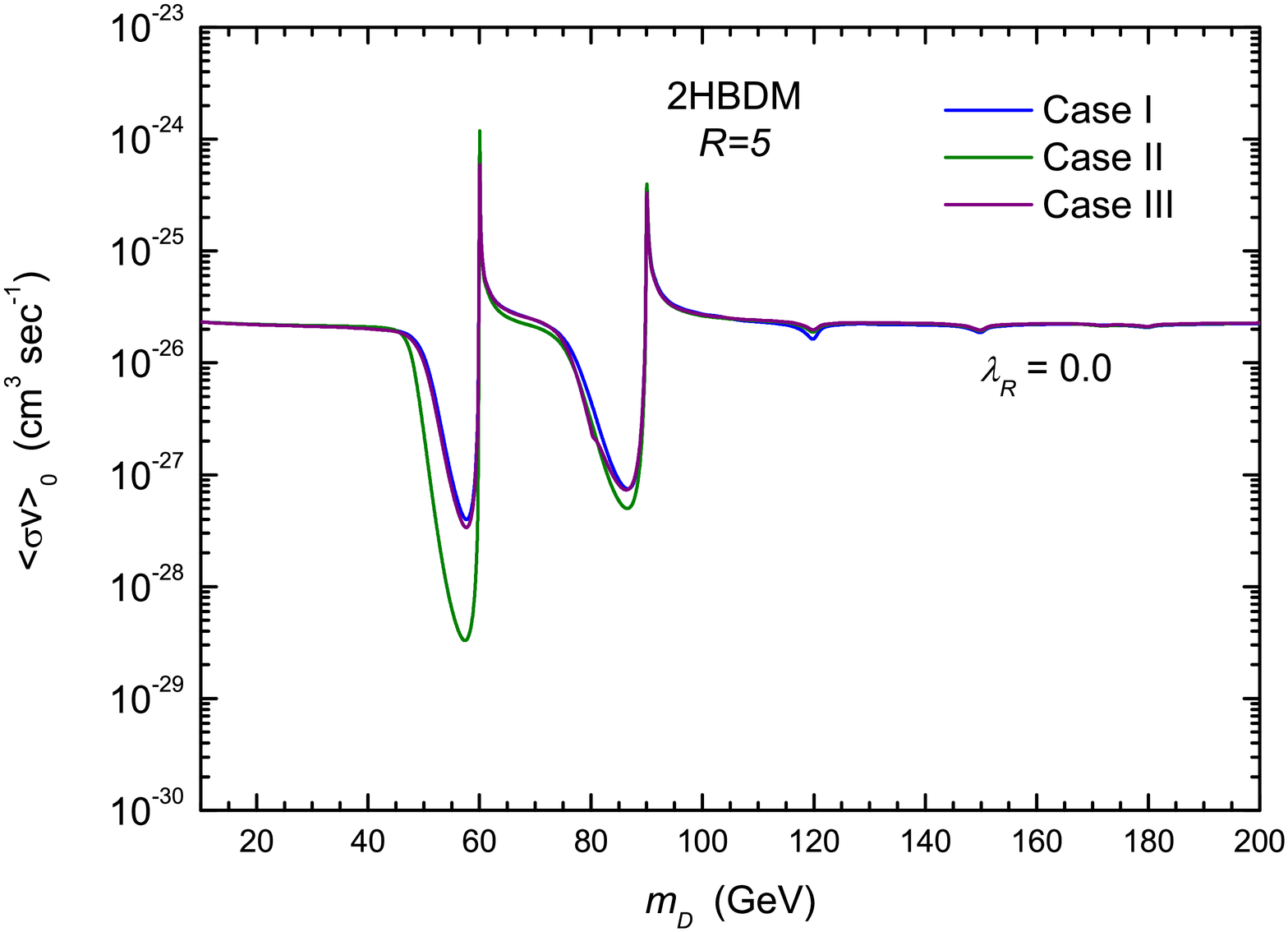}
\end{center}
\caption{  The predicted thermally averaged DM annihilation cross
section $\langle \sigma v \rangle_0$ in the 2HBDM. }
\label{Indirect2}
\end{figure}

For the DM indirect search, the 2HBDM has two enhanced regions for
$\langle \sigma v \rangle_0$ as shown in Fig. \ref{Indirect2}.
Therefore the PAMELA antiproton measurements can exclude two very
narrow regions. The predicted $\langle \sigma v \rangle_0$ is the
same as that in the 1HBDM for most parts of parameter space. When $
200 \,{\rm GeV} \leq m_D \leq 500\, {\rm GeV}$, one can still obtain
$\langle \sigma v \rangle_0 \approx 2.3 \times 10^{-26} \;{\rm cm}^3
\; {\rm sec}^{-1}$. It is clear that different mixing cases and $R$
scenarios lead to the same $\langle \sigma v \rangle_0$ except the
resonance regions.

In the 2HBDM, the DM-quark coupling $a_q$ in Eq. (\ref{fn}) is given
by
\begin{eqnarray}
a_q =  \frac{\lambda_{1,D} \, m_q }{2  m_D } \left ( \frac{f_1
}{m^2_{h}} + \frac{f_3 }{m^2_{H}}+ \frac{f_5 }{m^2_{A}} \right ) \,,
\label{aq2}
\end{eqnarray}
where $f_i$ have been presented in Appendix Eq. (\ref{f1-6}). Notice
that we have neglected the parameters $f_2$, $f_4$ and $f_6$ since
their contributions to $\sigma_{n}^{SI}$ are velocity-dependent.
Using the predicted $\lambda_{1, D}$ in Fig. \ref{L2HBDM} (left
panels), we calculate the spin-independent DM-nucleon elastic
scattering cross section $\sigma_{n}^{SI}$ for three mixing cases
and two $R$ scenarios. Different from  $\langle \sigma v \rangle_0$,
the predicted $\sigma_{n}^{SI}$ obviously depends on the mixing and
$R$ as shown in Fig. \ref{L2HBDM} (right panels). Although three
kinds of mixing cases have the almost same coupling $\lambda_{1,D}$
for $m_D \gtrsim 120$ GeV in the $R=1$ scenario, the predicted
$\sigma_{n}^{SI}$ in the Case III is far less than that in the Case
I and Case II. This is because that there is cancellation between
$f_1/m_h^2$ and $f_5/m_A^2$ in Eq. (\ref{aq2}) for the Case III.
When the DM candidate can annihilate into two light Higgs bosons, a
large $R$ does not obviously affect the predicted coupling
$\lambda_{1,D}$. However, the parameters $f_1, f_3$ and $f_5$ in Eq.
(\ref{aq2}) will be significantly enlarged. Therefore
$\sigma_{n}^{SI}$ usually increases as $R$ increases. The Case I
clearly demonstrates this feature. The enlarged $\sigma_{n}^{SI}$ in
the $R=5$ scenario may approach the CDMS II upper bound, which can
be used to explain the two possible events observed by the CDMS II
\cite{CDMSII}.  It is worthwhile to stress that the Case II in the
$R=5$ scenario give a smaller $\sigma_{n}^{SI}$ than that in the
$R=1$ scenario due to the cancellation from the different Higgs
boson contributions. We conclude that the predicted
$\sigma_{n}^{SI}$ is significantly dependent on the light Higgs
mixing and the parameter $R$. For $200 \,{\rm GeV} \leq m_D \leq
500\, {\rm GeV}$, the same conclusion can also be derived as shown
in Figs. \ref{R1} and \ref{R5} (right panels).

As shown in Figs. \ref{L2HBDM}, \ref{R1} and \ref{R5}  (right
panels), the CDMS II \cite{CDMSII} and XENON10 \cite{XENON10}
experiments can exclude the region $m_D \lesssim 50$ GeV. For $200
\, {\rm GeV} \leq m_D \leq 500$ GeV, our results show an upper bound
for $\sigma_{n}^{SI}$ which is still below the current experiment
upper bounds. The future experiments XENON100 \cite{XENON100P}, CDMS
100 kg \cite{CDMS100} and XENON1T \cite{XENON1T} can cover most
parts of the allowed parameter space except the extreme cancellation
cases. Nevertheless, it is still difficult to detect the DM direct
or indirect signals for the resonance regions $50\, {\rm GeV}
\lesssim m_D \lesssim 60$ GeV and $80\, {\rm GeV} \lesssim m_D
\lesssim 90$ GeV.

\section{Conclusions} \label{Conclusion}

In conclusion, we have investigated a scalar boson $S_D$ as the DM
candidate in the left-right symmetric gauge model with two Higgs
bidoublets, in which the SCPV can be easily realized. The stability
of DM candidate $S_D$ is ensured by the fundamental symmetries $P$
and $CP$ of quantum field theory. In order to well understand the DM
properties in the 2HBDM, we have firstly analyzed the 1HBDM and
shown that the predicted DM direct and indirect detection cross
sections ($\sigma_{n}^{SI}$ and $\langle \sigma v \rangle_0$) are
the same as that in the minimal extension of SM with a real singlet
scalar if $m_D < m_{h^0}$. When the annihilation channel $S_D S_D
\rightarrow h^0 h^0$ is open ($m_D > m_{h^0}$), the $H_2^0$ exchange
diagram relevant to $\lambda_R$ leads to a continuous DM-nucleon
elastic scattering cross sections $\sigma_{n}^{SI}$. Comparing with
the 1HBDM, there are more scalar particles entering the DM
annihilation and scattering processes in the 2HBDM. In the explicit
calculations, we have considered three typical mixing cases and two
Yukawa coupling scenarios ($R=1$ and $R=5$) to analyze the 2HBDM. It
has been shown that $\langle \sigma v \rangle_0$ is not sensitive to
the light Higgs mixing and Yukawa couplings except the resonance
regions. However $\sigma_{n}^{SI}$ is significantly dependent on the
above two factors. In general, $\sigma_{n}^{SI}$ can be enhanced by
large Yukawa couplings and approach the CDMS II upper bound, which
can be used to explain the two possible events observed by CDMS II.
It should be mentioned that a large Yukawa coupling may lead to a
very small $\sigma_{n}^{SI}$ in the extreme mixing case. Our results
show that the future DM direct search experiments can cover most
parts of the allowed parameter space. The PAMELA antiproton data can
exclude two very narrow regions in the 2HBDM. In addition, we have
shown that it is very difficult to detect the DM direct or indirect
signals for the resonance regions since the Breit-Wigner resonance
effect simultaneously suppresses $\sigma_{n}^{SI}$ and $\langle
\sigma v \rangle_0$.

\acknowledgments

This work is  supported in part by the National Basic Research
Program of China (973 Program) under Grants No. 2010CB833000;  the
National Nature Science Foundation of China (NSFC) under Grants No.
10975170, No. 10821504 and No. 10905084; and the Project of
Knowledge Innovation Program (PKIP) of the Chinese Academy of
Science.

\appendix

\section{Annihilation cross section }\label{Appendix.A}

For the annihilation processes $ S_D \, S_D \rightarrow f {\bar f}$,
the annihilation cross section $\hat{\sigma}_{f \bar f}$ is given by
\begin{eqnarray}
\hat{\sigma}_{f \bar f} = \sum_f m_f^2 \frac{\lambda_{1,D}^2 }{4
\pi} \sqrt{1- \frac{4m_f^2}{s}} \left[ (s-4m_f^2) P_1 + s P_2
\right] ,
\end{eqnarray}
where
\begin{eqnarray}
P_{1,2} = \left | \frac{f_{1,2}}{s- m_h^2 + i m_h \Gamma_h} +
\frac{f_{3,4}}{s- m_H^2 + i m_H \Gamma_H} +\frac{f_{5,6}}{s- m_A^2 +
i m_A \Gamma_A} \right|^2 \; ,
\end{eqnarray}
with
\begin{eqnarray}
& f_1  = c_x c_z - R c_y s_x - R c_x s_y s_z \;, & f_2 = R s_x s_y -
R c_x c_y s_z \;, \nonumber \\  & f_3= R c_x c_y +c_z s_x -R s_x s_y
s_z \;, & f_4 = -R s_x s_z c_y -R c_x s_y \;,\nonumber \\  & f_5= R
s_y c_z +s_z \;, \;\;\;\;\;\;\;\;\;\;\;\;\;\;\;\;\;\; & f_6 = R c_y
c_z \;. \label{f1-6}
\end{eqnarray}
The parameter $R$ has been defined in Eq. (\ref{R2}). The decay
widths of three light neutral Higgs are given by
\begin{eqnarray}
\Gamma_{h,H,A} = \frac{\sum m_f^2 }{8 \pi v_{\rm EW}^2} m_{h,H,A}
(f_{1,3,5}^2 + f_{2,4,6}^2) + \Gamma_{h,H,A}^{Z_1}\gamma_{h,H,A}  +
\Gamma_{h,H,A}^{W_1} \gamma_{h,H,A}  + \frac{\lambda_{1,D}^2 v_{\rm
EW}^2}{32 \pi} \frac{\sqrt{m_{h,H,A}^2 - 4 m_D^2}}{m_{h,H,A}^2} \;,
\end{eqnarray}
where $\gamma_{h} = c_x^2 c_z^2 $, $\gamma_{H} = s_x^2 c_z^2 $ and
$\gamma_{A} = s_z^2 $.  $\Gamma_{h,H,A}^{Z_1}$ and
$\Gamma_{h,H,A}^{W_1}$ have the following forms:
\begin{eqnarray}
\Gamma_{h,H,A}^{Z_1} & = & \frac{m_{h,H,A}^3}{32 \pi v_{\rm EW}^2}
\sqrt{1- \frac{4 m_{Z_1}^2}{m_{h,H,A}^2}} \left(1-
\frac{4m_{Z_1}^2}{m_{h,H,A}^2}+ \frac{12 m_{Z_1}^4}{m_{h,H,A}^4}\right)\;, \\
\nonumber \Gamma_{h,H,A}^{W_1} & = &  \frac{m_{h,H,A}^3}{16 \pi
v_{\rm EW}^2} \sqrt{1- \frac{4 m_{W_1}^2}{m_{h,H,A}^2}} \left(1-
\frac{4m_{W_1}^2}{m_{h,H,A}^2}+ \frac{12
m_{W_1}^4}{m_{h,H,A}^4}\right) \;.
\end{eqnarray}

For the annihilation processes $ S_D \, S_D \rightarrow Z_1 Z_1$ and
$ S_D \, S_D \rightarrow W_1 W_1$, we have
\begin{eqnarray}
\hat{\sigma}_{Z_1 Z_1} &=& \frac{\lambda_{1,D}^2 }{16 \pi}
\sqrt{1-\frac{4 m_{Z_1}^2}{s}} \left(1- \frac{4m_{Z_1}^2}{s}+
\frac{12 m_{Z_1}^4}{s^2} \right) \frac{s^2}{4} \nonumber \\ & \times
& \left| \frac{2 c_x c_z}{s- m_h^2 + i m_h \Gamma_h} + \frac{2 s_x
c_z}{s- m_H^2 + i m_H
\Gamma_H} +\frac{2 s_z}{s- m_A^2 + i m_A \Gamma_A} \right|^2 , \\
\hat{\sigma}_{W_1 W_1} &=& \frac{\lambda_{1,D}^2 }{8 \pi}
\sqrt{1-\frac{4 m_{W_1}^2}{s}} \left(1- \frac{4m_{W_1}^2}{s}+
\frac{12 m_{W_1}^4}{s^2} \right) \frac{s^2}{4} \nonumber \\ & \times
& \left| \frac{2 c_x c_z}{s- m_h^2 + i m_h \Gamma_h} + \frac{2 s_x
c_z}{s- m_H^2 + i m_H \Gamma_H} +\frac{2 s_z}{s- m_A^2 + i m_A
\Gamma_A} \right|^2 .
\end{eqnarray}

If the annihilation productions are a Higgs and a gauge boson, we
can derive
\begin{eqnarray}
\hat{\sigma}_{Z_1 A} &=&  \frac{\lambda_{1,D}^2 }{32 \pi}
\frac{\left[ (s-m_A^2 -m_{Z_1}^2)^2 - 4 m_A^2 m_{Z_1}^2
\right]^{1.5}}{s} \left| \frac{2 c_z s_x}{s- m_h^2 } - \frac{2 c_x
c_z}{s- m_H^2 }
\right|^2 \;, \nonumber \\
\hat{\sigma}_{Z_1 H}  &=& \frac{\lambda_{1,D}^2 }{32 \pi}
\frac{\left[ (s-m_H^2 -m_{Z_1}^2)^2 - 4 m_H^2 m_{Z_1}^2
\right]^{1.5}}{s} \left| \frac{2 c_x c_z}{s- m_A^2} -
\frac{2 s_z}{s- m_h^2 } \right|^2 \;, \nonumber \\
\hat{\sigma}_{Z_1 h} &=& \frac{\lambda_{1,D}^2 }{32 \pi}
\frac{\left[ (s-m_h^2 -m_{Z_1}^2)^2 - 4 m_h^2 m_{Z_1}^2
\right]^{1.5}}{s} \left| \frac{2 s_z}{s- m_H^2} - \frac{2 c_z
s_x}{s- m_A^2} \right|^2
\;, \nonumber \\
\hat{\sigma}_{W^\pm H^\mp} &=&  \frac{\lambda_{1,D}^2 }{4 \pi}
\frac{\left[ (s-m_{H^\pm}^2 -m_{W_1}^2)^2 - 4 m_{H^\pm}^2 m_{W_1}^2
\right]^{1.5}}{s} \left| \frac{a_1}{s- m_A^2 } + \frac{a_2}{s- m_H^2
}+ \frac{a_3}{s- m_h^2 } \right|^2,
\end{eqnarray}
where
\begin{eqnarray}
a_1 & = & c_y c_z - i c_z s_y \;, \nonumber \\ a_2  & = &  - c_x(i
c_y +s_y)-c_y s_x s_z+ i s_x s_y s_z \;, \nonumber \\ a_3  & = &  i
c_y s_x + s_y s_x - c_x (c_y s_z - i s_y s_z)\;.
\end{eqnarray}

When two DM candidates annihilate into two Higgs particles, we can
obtain
\begin{eqnarray}
  \hat{\sigma}_{k k} &=& \frac{\lambda_{1,D}^2 }{16 \pi}
\sqrt{1-\frac{4 m_k^2}{s}} \left[ G_2^2 - \frac{8 \lambda_{1,D}
v_{\rm EW}^2}{s-2 m_k^2} G_2 F({\xi_{kk}}) + \frac{8 \lambda_{1,D}^2
v_{\rm EW}^4}{(s-2 m_k^2)^2} \left(
\frac{1}{1-\xi_{kk}^2} + F({\xi_{kk}}) \right) \right]\;, \nonumber \\
  \hat{\sigma}_{i j} &=& \frac{\lambda_{1,D}^2 }{8
\pi} \beta_{ij}  \left[ G_2^2 - \frac{8 \lambda_{1,D} v_{\rm
EW}^2}{s- m_i^2 - m_j^2} G_2 F({\xi_{i j}}) + \frac{8
\lambda_{1,D}^2 v_{\rm EW}^4}{(s- m_i^2- m_j^2)^2} \left(
\frac{1}{1-\xi_{i j}^2} + F({\xi_{i j}})
\right) \right], \nonumber \\
 \hat{\sigma}_{H^\pm H^\mp}  &=& \frac{\lambda_{1,D}^2 }{8 \pi} \sqrt{1-\frac{4
m_{H^\pm}^2}{s}} G_2^2,
\end{eqnarray}
with
\begin{eqnarray}
G_2 &=& 1 + \frac{3 m_h^2}{s - m_h^2} + \frac{3 m_h^2}{s - m_H^2}+
\frac{3 m_h^2}{s - m_A^2} + \frac{\alpha_1 \lambda_{3,D} v_R^2}{s -
m_{H_2^0}^2} \frac{1}{\lambda_{1,D}} + \frac{m_\sigma^2}{s-
m_\sigma^2}.
\end{eqnarray}
The subscripts $k$ and $ij$ run over ($h, H, A$) and ($hH, hA, HA$),
respectively. The quantity $F$ is defined as
$F(\xi)\equiv\mbox{arctanh}(\xi)/\xi$ with $\xi_{i j} = \sqrt{1-4
m_D^2/s} \sqrt{(s- m_i^2- m_j^2)^2-4 m_i^2 m_j^2}/ (s- m_i^2-
m_j^2)$. The parameter $\beta_{ij}$ is given by $\beta_{ij} =
\sqrt{(s- m_i^2- m_j^2)^2-4 m_i^2 m_j^2}/{s}$.



\begin{thebibliography}{99}
\bibitem{modify} Here we have replaced the incorrect expression of Eq. (15) in
Ref. [6], which only affects the numerical results and does not
change the conclusions.
\end{thebibliography}
\end{document}